\documentclass[%
 reprint,
 superscriptaddress,
 nofootinbib,
 amsmath,amssymb,
 aps,
 prd,
floatfix,
]{revtex4-2}

\usepackage{placeins}
\usepackage{graphicx}
\usepackage{dcolumn}
\usepackage{bm}
\usepackage{color, xcolor}
\usepackage{appendix}

\usepackage[normalem]{ulem}

\newcommand{\fmin}[0]{f_\textrm{min}}
\newcommand{\fmax}[0]{f_\textrm{max}}
\newcommand{\Chat}[0]{\hat C(f)}

\newcommand{\boldm}[0]{\mathbf{m}}
\newcommand{\boldd}[0]{\mathbf{d}}
\newcommand{\bolda}[0]{\mathbf{a}}
\newcommand{\boldx}[0]{\mathbf{x}}

\newcommand{\Omegagw}[0]{\Omega_{\textrm{GW}}}

\begin{document}

\preprint{APS/123-QED}

\title{Model-agnostic gravitational-wave background characterization algorithm}

\author{Taylor Knapp}
\email{tknapp@caltech.edu}
\affiliation{Theoretical Astrophysics Group, California Institute of Technology, Pasadena, California 91125, USA}
\affiliation{LIGO Laboratory, California Institute of Technology, Pasadena, California 91125, USA}

\author{Patrick M. Meyers}
\email{pmeyers@caltech.edu}
\affiliation{Theoretical Astrophysics Group, California Institute of Technology, Pasadena, California 91125, USA}

\author{Arianna I. Renzini}
\email{arianna.renzini@unimib.it}
\affiliation{Dipartimento di Fisica “G. Occhialini”, Universit\'a degli Studi di Milano-Bicocca, Piazza della Scienza 3, 20126 Milano, Italy}
\affiliation{INFN, Sezione di Milano-Bicocca, Piazza della Scienza 3, 20126 Milano, Italy}

\date{\today}

\begin{abstract}
As ground-based gravitational-wave (GW) detectors improve in sensitivity, gravitational-wave background (GWB) signals will progressively become detectable. 
Currently, searches for the GWB model the signal as a power law; however, deviations from this model will be relevant at increased sensitivity. 
Therefore, to prepare for the range of potentially detectable GWB signals, we propose an interpolation model implemented through a transdimensional reversible-jump Markov chain Monte Carlo algorithm. %
This interpolation model foregoes a specific physics-informed model (of which there are a great many) in favor of a flexible model that can accurately recover a broad range of potential signals.
In this paper, we employ this framework for an array of GWB applications. %
We present three dimensionless fractional GW energy density injections and recoveries as examples of the capabilities of this spline interpolation model. %
We further demonstrate how our model can be implemented for hierarchical GW analysis on $\Omegagw$.

\end{abstract}

\maketitle

\section{\label{sec:introduction}Introduction}
Gravitational-wave (GW) detection is rapidly evolving into a precision science involving a variety of measurement techniques and detection strategies. 
In contrast to the individual compact binary signals observed with ground-based GW detectors thus far~\cite{Abbott_2023}, the gravitational-wave background (GWB) is the superposition of all subthreshold signals that cannot be individually resolved by the detector network. 
Recently, multiple pulsar timing array (PTA) collaborations presented evidence for a GWB in the nanohertz frequency band~\cite{NANOGrav:2023gor, EPTA:2023fyk, PPTA, CPTA}. That signal is consistent with the predictions for a population of supermassive black hole binaries~\cite{agazieNANOGrav15Yr2023a}.
No gravitational-wave background has been detected in the $10-1000$ Hz band, where detectors such as Advanced LIGO~\cite{LIGOScientific:2014pky}, Advanced Virgo~\cite{Acernese2015}, and KAGRA operate~\cite{KAGRA_design_doc}; however, such a discovery could provide a wealth of information.

There is no shortage of predicted GWB signals in the ground-based detector band. On the astrophysical side, a signal from unresolved compact binary coalescences (CBCs) carries information about populations of binary black holes (BBHs)~\cite{PhysRevX.13.011048}, high redshift metallicity~\cite{fishbach2021}, and populations of black holes in globular clusters~\cite{Romero-Shaw:2020siz,Kou:2024gvp}. 
On the cosmological side, signals from first-order phase transitions~\cite{Grischuk1975,Hogan:1986dsh,Huber:2008hg,Child:2012qg,Hindmarsh:2013xza,Hindmarsh:2015qta,Hindmarsh:2017gnf,Jinno:2016vai,Weir:2017wfa}, cosmic strings~\cite{Battye1997,vilenkinCosmicStringsOther2000,Jeannerot:2003qv,Damour:2004kw}, or inflationary models~\cite{Guzzetti:2016mkm}, among many others, have also been proposed~\cite{Caprini2018}. 
Generally, the CBC signal is expected to dominate the band, and predictions show that detection is possible with improvements to the current detector network~\cite{KAGRA:2021kbb}.

The canonical model for the GWB spectrum from $10-1000$ Hz is a power law (PL), and there are two primary reasons for adopting this model. 
Firstly, current interferometers (often referred to as second-generation --2G-- detectors) are not sufficiently sensitive to GWs,~\footnote{Currently only very loud and/or nearby individual signals can be detected, which may be viewed as outliers of the total GW emitter population. The sensitivity to subthreshold background signals is therefore low; this is further compounded by the fact that the GWB is usually targeted using a search for persistent isotropic background signals~\cite{KAGRA:2021kbb}, which is systematically less sensitive to a CBC background in this frequency band, as this signal is neither persistent~\cite{Lawrence:2023buo} nor isotropic~\cite{Jenkins_2018}.} so an initial detection of the GWB will likely be possible using a simple model that reflects that sensitivity. 
Secondly, most models for a GWB predict a (approximate) PL shape in the ground-based detector band. For the unresolved CBC background, for example, it is predicted that most of the GW power in the sensitive frequency band comes from the early inspiral phase of compact binaries, with a gradual turnover in the spectrum at higher frequencies, where the detectors are less sensitive. However, there are several reasons to expect deviations from this typical PL assumption: these include sensitivity improvements in upcoming detectors~\cite{Cutler:2005qq,Sachdev:2020bkk,Sharma:2020btq}, deviations due to specific population models or waveform choices~\cite{Renzini:2024pxt}, distinctive features in cosmological stochastic signals~\cite{Grischuk1975,Hogan:1986dsh,Huber:2008hg,Child:2012qg,Hindmarsh:2013xza,Hindmarsh:2015qta,Hindmarsh:2017gnf,Jinno:2016vai,Weir:2017wfa,Martinovic:2020hru}, and the possibility of distinguishing multiple correlated signals~\cite{Meyers:2020qrb,janssensImpactSchumannResonances2021,janssensCorrelated11000Hz2023}. 

The detection and characterization of the GWB spectrum can be leveraged to infer properties of the stellar-mass black hole population. 
This includes, for example, probing the merger rate as a function of cosmological redshift~\cite{Callister:2020arv,Renzini:2024pxt}, typically using a Madau-Dickinson-like merger rate evolution~\cite{PhysRevX.13.011048, KAGRA:2021kbb}. 
One expects deviations from this fiducial model through potential subpopulations of primordial black holes as well as uncertainties in the star formation rate at high redshifts (see Sec.~\ref{ssec:case_for_flexible_model}).

In this paper, we outline the scientific case for going beyond the power law assumption for GWBs and the Madau-Dickinson-like redshift evolution, and we introduce a framework for detecting and characterizing the GWB or the BBH merger rate evolution based on the reversible jump Markov chain Monte Carlo (RJMCMC) algorithm~\cite{greenReversibleJumpMarkov1995}. For modeling the GWB spectrum, for example, we interpolate a set of ``knots'' arranged in frequency space to the frequency bins of the measured data. The amplitude of the GWB and the location in the frequency space of each knot, as well as the \textit{number} of knots, are independent parameters in the model. By making the number of knots variable, the interpolation model can fit a variety of spectral shapes and can yield a detection statistic for a GWB.  For the merger rate as a function of redshift, we use the same flexible approach, instead placing the knots in redshift space.

We focus on ground-based detector spectral ranges and use Advanced LIGO, Advanced Virgo, and Cosmic Explorer~\cite{Evans:2021gyd, Evans:2023euw} design sensitivities in our data analysis examples. The methods can be applied to other frequency bands and detection techniques and will have valuable applications for both PTAs and the upcoming Laser Interferometer Space Antenna (LISA). 

The rest of this paper is organized as follows. In Sec.~\ref{sec:gwb} we discuss current models for the GWB, including astrophysical models, and make the case for using a more flexible model. In Sec.~\ref{sec:search_methods_overview} we review search methods for a GW background, while in 
Sec.~\ref{sec:rjmcmc} we discuss the spline interpolation model for the GWB and the transdimensional fitting algorithm.
In Sec.~\ref{sec:fit_omgw} we apply our new spectral models to two realistic examples for the CBC-generated GWB and to a generic broken power law (BPL) background. We show results on simulated data for both current and next-generation detectors.
In Sec.~\ref{sec:constrain_Rz} we demonstrate how to use the new framework to fit the BBH merger rate as a function of redshift using only measurements from the GWB searches.
In Sec.~\ref{sec:disc_and_conc} we discuss current and future applications of this method and offer concluding remarks.

\section{The GWB signal}
\label{sec:gwb}
The GWB signal is typically characterized in terms of the dimensionless fractional GW energy density per logarithmic frequency bin:
\begin{align}
    \Omegagw(f) &= \frac{1}{\rho_c}\frac{d\rho_{\rm gw}}{d\log f},
\end{align}
where $\rho_{\rm gw}$ is the energy density in gravitational waves in the Universe,  $\rho_c=3c^2H_0^2 / 8\pi G$ is the critical energy density to close the Universe and $H_0$ is the Hubble rate today. We use $H_0=67.66$ km/s/Mpc from the \textit{Planck} satellite 2018 results~\cite{Planck:2018vyg}. In some papers the astrophysical GWB is characterized in terms of the GW flux (e.g.~\cite{KAGRA:2021mth,Bellie:2023jlq}), but we choose to express the signal in terms of $\Omegagw$ here for comparison with past results from ground-based experiments.

In the rest of this section, we first discuss current models that are used for $\Omegagw(f)$, as well as $R(z)$, the evolution of the merger rate of black holes. We then discuss potential deviations from those fiducial models. 

\subsection{Current models for the GWB spectrum}
\label{ssec:gwb_signal:current_models}
If the dominant contribution to the GWB comes from the early inspiral phase of CBCs, the spectrum is expected to be a PL in the frequency band $\mathcal{O}(10–100 \, \textrm{Hz})$, where current ground-based detectors are most sensitive:
\begin{align}
\Omegagw(f) = \Omega_{\textrm{ref}}\left(\frac{f}{f_{\textrm{ref}}}\right)^{\alpha},
\end{align}
with spectral index $\alpha = 2/3$ and reference amplitude $\Omega_{\textrm{ref}}$, which depends upon the population of systems that produce the signal, and is typically left as a free parameter. This prediction is based on waveforms computed using a zeroth-order post-Newtonian expansion~\cite{Ajith:2007kx}, under the assumption that each binary is far from merger.

Other GWB models, such as those arising from slow-roll inflation~\cite{Caprini2018} or cosmic strings~\cite{Battye1997,vilenkinCosmicStringsOther2000,Jeannerot:2003qv,Damour:2004kw}, also predict PL spectra, but each with a different spectral index. To accommodate the range of potential PL backgrounds, the joint posterior distribution $p[\Omega_{\textrm{ref}}, \alpha|\Chat]$ is usually fit using Markov chain Monte Carlo (MCMC) methods, where $\Chat$, described in detail in Sec.~\ref{sec:search_methods_overview}, is the measured (normalized) cross-correlation spectrum between detectors. More complex models are also considered, including BPL spectra to approximate features from first-order phase transitions in the early Universe~\cite{Grischuk1975,Hogan:1986dsh,Huber:2008hg,Child:2012qg,Hindmarsh:2013xza,Hindmarsh:2015qta,Hindmarsh:2017gnf,Jinno:2016vai,Weir:2017wfa}, and models that change the overlap reduction function due to the inclusion of nontensorial polarizations~\cite{Callister:2017ocg, Tsukada:2023stv}.

Recent computational advances have enabled population-based modeling of the GWB and the simultaneous fitting of individual events and the unresolved background~\cite{smithOptimalSearchAstrophysical2018,Callister:2020arv,KAGRA:2021duu,KAGRA:2021kbb,Turbang:2023tjk, Lalleman:2025xcs}. In this approach, the unresolved (or unsubtracted) CBC population is fit directly to the cross-correlation spectrum $\hat{C}(f)$, with parametrized priors on masses, spins, and redshifts.
A “reference bank” of waveforms allows rapid adjustments to a fiducial spectral model based on changes in population assumptions~\cite{Turbang:2023tjk,Renzini:2024pxt}.
Although these models are powerful, their performance is highly dependent on prior choices and population assumptions, complicating their interpretation.

Using these techniques, one can fit the merger rate as a function of cosmological redshift~\cite{Callister:2020arv,Renzini:2024pxt}, typically assuming that binaries form in the field so the merger rate follows the star formation rate convolved with some time delay between formation and merger. This leads to a Madau-Dickinson-like merger rate evolution~\cite{PhysRevX.13.011048, KAGRA:2021kbb}. One can also potentially constrain the metallicity dependence on the evolution of BBHs~\cite{Turbang:2023tjk}, and probe the contribution to the GWB from hierarchical mergers~\cite{Kou:2024gvp}.

\subsection{The case for a flexible model}\label{ssec:case_for_flexible_model}
While PLs are convenient models for $\Omegagw(f)$, and well-founded in the lower SNR regime, several factors could lead to deviations from a PL. Additionally, when fitting the population properties of BBHs through their effect on $\Omegagw(f)$, the assumptions that are often made for, e.g. the redshift evolution of the merger rate, can also potentially break down.

When the background is dominated by CBCs, the frequency at which the spectrum turns over and the deviations from the PL shape depend on the distribution of masses of the merging binaries, along with the redshift at which they merge. 
Populations of compact binaries merging at different frequencies produce unique imprints in the overall background spectrum -- for example, the GWB due to a population of heavy BBHs at high redshifts will have a turnover at lower frequencies than a population of binary neutron star (BNS) or neutron star-black hole mergers. These features can be probed directly in background searches. Furthermore, if these populations have roughly comparable contributions at lower frequencies (e.g. $\sim 10-100$ Hz), then the turnover in the spectrum due to one population could distort the shape of the combined spectrum from all sources, requiring joint modeling. 

Furthermore, recent studies that use more realistic waveforms to make GWB predictions challenge the simple PL assumption at low frequencies because the contribution of the \textit{late inspiral} phase will change the shape of the spectrum -- especially the specific value of the spectral index~\cite{Renzini:2024pxt}. However, detecting such a deviation will likely require next-generation detectors. In this context, as detectors become more sensitive and resolvable CBC signals are subtracted before searching for a GWB, the residual GWB after subtraction may no longer take the same expected PL shape~\cite{Cutler:2005qq,Sachdev:2020bkk,Sharma:2020btq}.

Looking beyond CBC backgrounds, models like first-order phase transitions in the early Universe predict a BPL shape, where the spectral indexes before and after the break correspond to physically predictable quantities and the frequency where the spectral index changes corresponds to the temperature of the Universe when the phase transition took place
\cite{Grischuk1975,Hogan:1986dsh,Huber:2008hg,Child:2012qg,Hindmarsh:2013xza,Hindmarsh:2015qta,Hindmarsh:2017gnf,Jinno:2016vai,Weir:2017wfa,Martinovic:2020hru}. Finally, correlated magnetic noise between detectors could have a complicated spectral shape that will need to be characterized and separated from a true GWB signal~\cite{Meyers:2020qrb,janssensImpactSchumannResonances2021,janssensCorrelated11000Hz2023}.

Next, we turn to the merger rate vs. redshift distribution for binary black holes that contribute a GW background. Deviations from the typical Madau-Dickinson-like model could come due to lensing of high-redshift black holes~\cite{Broadhurst:2018saj,Broadhurst:2020moy,Broadhurst:2022tjm}, excess star formation at high redshifts hinted at in recent James Webb Space Telescope results~\cite{finkelsteinCEERSKeyPaper2023,harikaneComprehensiveStudyGalaxies2023}, or the effects of metallicity on the formation of black holes at high redshift~\cite{lehoucqAstrophysicalUncertaintiesGravitationalwave2023,perigoisStartrackPredictionsStochastic2021,kowalska-leszczynskaEffectMetallicityGravitationalwave2015,martinovicFootprintsPopulationIII2022,nakazatoGRAVITATIONALWAVEBACKGROUND2016,dvorkinMetallicityconstrainedMergerRates2016,Turbang:2023tjk}.

All of these added complexities, which can potentially be probed with next-generation detectors, motivate a more flexible model than a simple PL. 
In this paper, we use linear and spline interpolation with a variable number of knots to flexibly model the GWB spectrum and the BBH merger rate as a function of redshift.
Although it may seem less informative compared to physical models used in parameter estimation, our model avoids the pitfalls of increased prior volume (and corresponding Occam penalties), reduces trials factors associated with testing multiple models, and could be used to fit correlated noise in the future as well. 

In subsequent sections, we will discuss how to interpret the results of this flexible fitting procedure in the context of established physical models, highlighting its utility both for detection and for guiding astrophysical inference.

\section{Search methods overview}
\label{sec:search_methods_overview}
In this section we outline the main methods used to search for a GW background with ground-based detectors--readers already familiar with this material can skip to Sec.~\ref{sec:rjmcmc}.

Common assumptions that simplify stochastic GW analyses are that the GWB is isotropic, broadband, stationary, Gaussian, and unpolarized~\cite{Allen:1999}. In this limit, a sufficient search strategy is to cross-correlate data from pairs of instruments. Although there are search strategies for detecting and characterizing a GWB that is non-Gaussian~\cite{Drasco:2002yd,smithOptimalSearchAstrophysical2018,Buscicchio:2022raf,Lawrence:2023buo}, anisotropic~\cite{Allen:1996gp,Ballmer:2005uw,Mitra:2007mc,Thrane:2009fp,Ain:2018zvo, Renzini:2018vkx, Tsukada:2022nsu}, narrow band~\cite{Ballmer:2005uw}, or some combination of each~\cite{Thrane:2015aua,Goncharov:2018ufi,KAGRA:2021rmt, Agarwal:2023lzz, Xiao:2022uvq}, we focus on the isotropic, Gaussian, unpolarized and stationary case.

A hybrid frequentist-Bayesian approach is used to perform the search~\cite{MandicThrane2012,Romano:2016dpx,Matas:2020roi, pygwb_paper}. 
First, an unbiased frequentist estimator for the strength of the background in individual frequency bins is constructed by cross-correlating many chunks of data~\cite{Allen:1999,Romano:2016dpx, pygwb_paper}:
\begin{align}
    \hat C_{ij}(f;t) = \frac{2}{T}\frac{\textrm{Re}\left[\tilde s_{i}^*(f; t) \tilde s_j(f; t)\right]}{\Gamma_{ij}(f)S_0(f)},
\end{align}
where $\tilde s_i(f; t)$ is the Fourier transform of the strain data in detector $i$, starting at time $t$ and ending at time $t+T$, with $T=192$ s in most recent analyses~\cite{KAGRA:2021kbb}. $\Gamma_{ij}(f)$ is the overlap reduction function~\cite{Christensen:1992wi,Flanagan:1993ix}, and $S_0(f)=3H_0^2 / (10 \pi^2 f^3)$ is a normalization factor that makes $\langle \hat C_{ij}(f)\rangle=\Omegagw (f)$ (in the absence of correlated noise~\cite{Meyers:2020qrb}). 

In the limit where the GW strain amplitude is much smaller than the typical noise in the detectors [on the short, $\mathcal O(100\,\textrm{s})$ timescales that $\hat C_{ij}(f;t)$ is calculated over], the variance of $\hat C_{ij}(f; t)$ is given by~\cite{Romano:2016dpx}:
\begin{align}
\label{eq:variance_on_omega_estimator}
    \sigma^2_{ij}(f;t) = \frac{1}{2\Delta f T}\frac{P_i(f;t)P_j(f;t)}{\Gamma_{ij}(f)^2S_0(f)^2},
\end{align}
where $\Delta f$ is the width of the frequency bins used in the analysis (0.3125 Hz in most recent LVK Collaboration results~\cite{KAGRA:2021kbb}) and $P_i(f; t)$ is the one-sided power spectral density (PSD) of the noise in detector $i$.  

The $\hat C_{ij}(f;t)$ spectra are combined across time and between detector pairs using an inverse noise-weighted average, leaving a final frequentist estimator for the GWB in each frequency bin $\hat C(f)$, with associated variance $\sigma^2(f)$. 

In situations where the detector noise does not dominate the signal, $\sigma^2(f)$  [or potentially even $\sigma_{ij}^2(f;t)$] would need to be estimated either independently or as a part of the Bayesian step we discuss below. This is likely to be the situation for third-generation detectors, where we may no longer be in the weak signal regime.  In this work, we do not account for the intermediate or large signal regime. We assume that we have an independent measurement of the detector noise PSDs. We discuss potential methods for handling the intermediate and strong signal regimes in Sec.~\ref{sec:disc_and_conc}. 

Next, the frequentist statistic is used to perform a weighted average across frequency bins assuming a specific spectral model~\cite{Allen:1999, Romano:2016dpx,pygwb_paper}, to estimate the background signal at a reference frequency.
The same spectral model that fits $\Chat$ for the spectral shape and amplitude of the background can also be used in Bayesian parameter estimation~\cite{MandicThrane2012,Callister:2017ocg,Meyers:2020qrb,pygwb_paper}.
The $\Chat$ spectrum is obtained by averaging many time segments together, and so the central limit theorem implies that it should be Gaussian distributed. Therefore, the likelihood used for a Bayesian analysis of $\Chat$ is given by~\cite{MandicThrane2012,Romano:2016dpx,Matas:2020roi}
\begin{widetext}
\begin{align}
\label{eq:ln_likelihood}
    \ln p(\hat C(f) | \bm\theta, \mathcal M) = -\frac{1}{2}\sum_f \left\{\frac{[\hat C(f) - \Omega_{\mathcal{M}}(f, \bm\theta)]^2}{\sigma^2(f)} + \ln(2 \pi \sigma^2(f))\right\},
\end{align}
\end{widetext}
where we use $\mathcal M$ to label the model, $\Omega_{\mathcal M}(f)$ the spectrum the model generates, and $\bm\theta$ the vector of parameters associated with the model. 

While we discussed the broad outline of the search method in this section, for the injection studies in Sec.~\ref{sec:fit_omgw} we do not simulate individual time segments or time domain data--instead we calculate $\sigma^2(f)$ using publicly available sensitivity curves, and simulate the final $\Chat$ using this variance.

\section{Flexible, Transdimensional Models}
\label{sec:rjmcmc}

We propose an interpolation model to recover a GWB signal by placing a discrete number of points in an $n$-dimensional space and interpolating between these points.
This requires a choice of interpolation scheme, which we describe in Sec. \ref{sec:intrp_models}. 
The general interpolation model naturally lends itself to using a RJMCMC algorithm~\cite{greenReversibleJumpMarkov1995} that allows the number of knots and thus the number of parameters in our model to vary. 
So, we must define how these interpolating points, or \textit{knots}, move for each proposal in the RJMCMC. 
These knot movements are our \emph{proposals} and are defined in Sec. \ref{sec:proposals}. 
This is similar in spirit to the BayesLine and BayesWave algorithms~\cite{Littenberg:2014oda,Cornish:2014kda,Cornish:2020dwh}. 

Similar interpolation schemes have been used to flexibly fit unknown spectral shapes in the past. For example, in~\cite{renziniImprovedLimitsStochastic2019,Xiao:2022uvq} the authors estimate maps of the GWB from LVK data independently in separate frequency bins, with a flexible binning scheme in~\cite{Xiao:2022uvq} based on the sensitivity to the GWB spectrum. A similar approach was presented in~\cite{capriniReconstructingSpectralShape2019a}, where the authors employ frequency binning across the LISA spectrum using bins chosen based on the expected sensitivity curve. In~\cite{ghalebBayesianReconstructionPrimordial2025} the authors use a similar interpolation scheme to what we present below, with a variable number of knots. They choose the number of knots by repeatedly calculating the Bayesian evidence, while in this work we choose the knots dynamically with the reversible jump algorithm. 

\subsection{Interpolation models} \label{sec:intrp_models}
Instead of using a specific functional form for the GWB, such as a PL, we instead estimate the GWB at a set of chosen frequencies, which we call ``knots,'' and interpolate from those knots to the frequency bins at which measurements are taken. 
The parameters of the model are a set of $n$ amplitudes $\bolda_n$, located at a set of knots $\boldx_n$, in frequency space. We use $x$ here to denote the location of a knot, to distinguish them from the frequency bins at which $\Chat$ is calculated. 

We consider a few interpolation methods. The first method performs linear interpolation in log-space from the knots to the frequency bins. For a measurement at frequency $f$, our model is 
\begin{align}
\label{eq:linear_interp}
    \mathcal G(f) &= c_j (f - x_j) + a_j\,,\\
    c_j &= \frac{a_{j+1} - a_j}{x_{j+1} - x_{j}}.
\end{align}
with $f \in [x_j, x_{j+1}]$. We refer to this model as the ``linear interpolation'' model or, if we perform the interpolation in log-space (i.e. $x\rightarrow \ln x$ and $a \rightarrow \ln a$ etc.) the ``piecewise power law'' (PPL) model because linear interpolation in log-space corresponds to constructing the model from a set of PPLs. By working in log-space we can ensure that the GWB is positive (a defining characteristic of $\Omegagw$), and that in the case of $n=1$ or $n=2$ knots the model simplifies to a constant or PL model. Moreover, for $n=2$, placing a log-uniform prior on $\bolda$ means that when we recover a PL, the prior we use is naturally the preferred prior advocated for in~\cite{Callister:2017ocg} and used in~\cite{Callister:2017ocg,KAGRA:2021kbb}. 

Another potential model is the ``cubic spline'' model, where we interpolate using cubic splines instead of performing linear interpolation. In that case, the model is 
\begin{align}
\label{eq:cubic_spline_interp}
\mathcal G(f) &= \sum_{n=0}^3\,c_{n,j}(f - x_j)^n,
\end{align}
where the $c_{n,j}$'s are calculated using standard algorithms that solve for the $c_{n,j}$ subject to constraints that impose continuity and smoothness~\cite{deboor}.

An additional interpolation method that we employ is the ``Akima spline'' model. 
Akima splines are nonsmoothing splines that avoid the rapid oscillations between knots that are common in other spline formulations when the data are uninformative or the second derivative of the underlying function varies rapidly. 
An Akima spline model takes the same form as the cubic spline, but the coefficients $c_{n,j}$ are informed by derivatives of neighboring knots instead of an optimization \cite{WANG2014122}. 
The result of an Akima spline interpolation is a smoother interpolation that is less sensitive to outliers in the data. 

\subsection{Reversible jump Markov chain Monte Carlo} \label{sec:proposals}
We start with an underlying grid of $N$ knots that are available to interpolate between. At any given time, $n$ of them will be used in the interpolation, or ``activated,'' and $N-n$ knots will be disregarded in the interpolation, or ``deactivated.''
The locations of the available underlying knots are labeled $x_\alpha$, where $\alpha=0, \ldots, N-1$. So, the initial location of $x_\alpha$ is given by 
\begin{align}
x_\alpha = \alpha\left(\frac{\fmax - \fmin}{N-1}\right) + \fmin.
\end{align}
We use $\alpha$ to index the underlying grid of \emph{potential} knots, and subscripts $j$ or $n$ to index the grid of \emph{activated} knots. We use $j$ to index individual knots, $x_j$, and $n$ will be used to indicate the size of the vector containing all activated knots $\boldx_n$.

The parameters of our model are the list of amplitudes $\bolda_{n}$ corresponding to the amplitude of the model at the knot location $\boldx_n$ of each of the activated knots. We also allow the knot locations to vary but maintain ordering on the underlying grid. That is, $x_j$ can vary within the range $\fmin + 2j\Delta x < x_j< \fmin+2(j+1)\Delta x$, where $\Delta x = (\fmax - \fmin)/(2N)$.
So at any given time we have $2n$ parameters in our model. The prior on $\boldx_n$ and $\bolda_n$ throughout this paper is uniform within their allowed regions, and disjoint between knots~\footnote{This simply implies two knots cannot be superposed.}:
\begin{align} \label{eq:priors}
    p(\bolda_n, \boldx_n) = (2\Delta x)^{-n}(a_{\textrm{max}} -a_{\textrm{min}})^{-n}. 
\end{align}
We discuss the effect of these priors and how they change as the number of knots changes in Appendix ~\ref{app:transdimensional}.

\begin{figure}[]
    \includegraphics[width=\columnwidth,clip=true]{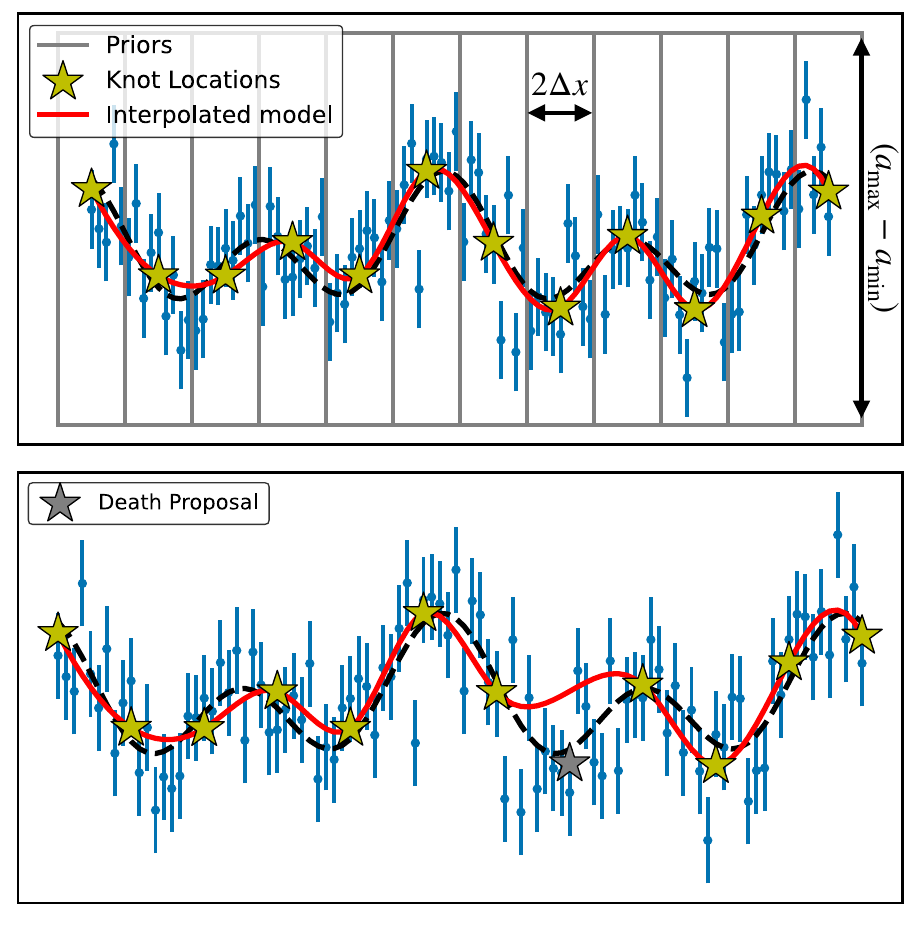}\\
     \caption{ \label{fig:schematic} 
     A visualization of the RJMCMC implemented in this work.
     In both panels, we show the injected signal (black dashed line) and the injected signal plus Gaussian white noise (blue points). 
     The error bars on the blue points correspond to one standard deviation of the Gaussian white noise. 
     In the top panel, the gray boxes correspond to the uniform priors that the interpolating knots (yellow stars) are allows to move within [see Eq.~\eqref{eq:priors}].
     The knots are interpolating using cubic splines (red lines).
     The bottom panel shows the death proposal (see Sec.~\ref{sec:proposals}) where we have turned off one knot (gray star) and interpolated the remaining knots to fit the signal.
     This schematic highlights the transdimensional nature of the RJMCMC that we implement to fit injected GW signals with various detector sensitivities. 
     }
\end{figure}

We use MCMC to evaluate posterior distributions on select parameters. At each step in the Markov chain, a new point is proposed and that new point is accepted according to some acceptance probability. In the reversible jump algorithm, the new point that is proposed can have either a different number of parameters compared to the current point or the same number of parameters. In the analyses below, we use four different jump proposals\footnote{We have tested the sampler and the jump proposals using prior recovery tests to ensure that it obeys detailed balance.}:
\begin{enumerate}
    \item \textit{Birth proposal}: We propose to ``turn on'' one knot that is currently deactivated, choosing from the list of deactivated knots at random. This increases the number of parameters in the model and the overall prior volume. The location and amplitude of the proposed knot is either drawn from the prior or the proposed $x_j^{new}$ is drawn from the prior, while $a_j$ is drawn from a Gaussian with a mean given by $\mathcal G(x_j^{new})$. In this case, the location and amplitudes of the knots for the interpolation, $x_j$ and $c_{n,j}$ in Eq.~\eqref{eq:linear_interp} or Eq.~\eqref{eq:cubic_spline_interp}, are from the current sample. The standard deviation of the Gaussian is selected by the user when the sampler is instantiated (the default is $\sigma=1$).
    \item \textit{Death proposal}: We propose to ``turn off'' a knot that is currently active. This decreases the number of parameters in the model and the overall prior volume.
    \item \textit{Shift knot horizontally}: We choose an active knot at random and propose to shift its location $x_j$ within its allowable region.
    \item \textit{Shift knot vertically}: We choose an active knot at random, and propose to shift its amplitude $a_j$, either drawing a point randomly from the prior, or drawing from a Gaussian centered on the current location. The scale of the Gaussian is chosen randomly from a uniform distribution between 0.1 and 3.
\end{enumerate}
Figure ~\ref{fig:schematic} provides a schematic of the RJMCMC, illustrating a death proposal and visualizing the priors associated with each interpolating knot.
We discuss the details of the algorithm in Appendix ~\ref{app:transdimensional}, including the basics of RJMCMC, and more details about the implementation of the jump proposals. 

\section{Detecting and Characterizing $\Omega_{\rm GW}(f)$}
\label{sec:fit_omgw}
We choose three example injections to demonstrate our transdimensional spline interpolation model. 
These injections include a BBH-induced signal (Sec. \ref{sec:CBC}), a subtraction signal (Sec. \ref{sec:subtraction}), and a signal based on a first-order phase transition model (Sec. \ref{sec:fopt}). 
For each example, after motivating the astrophysical relevance of each injection, we present the results of our injection and recovery. 
We perform the injection and recovery with both LIGO A+ and projected Cosmic Explorer (CE) sensitivity curves. For the CE sensitivity curve, we assume 40 km detector arms and the same location and orientation as the current LIGO interferometers at Hanford, WA and Livingston, LA \cite{Evans:2021gyd, Evans:2023euw}.

\begin{figure}[]
    \includegraphics[width=\columnwidth,clip=true]{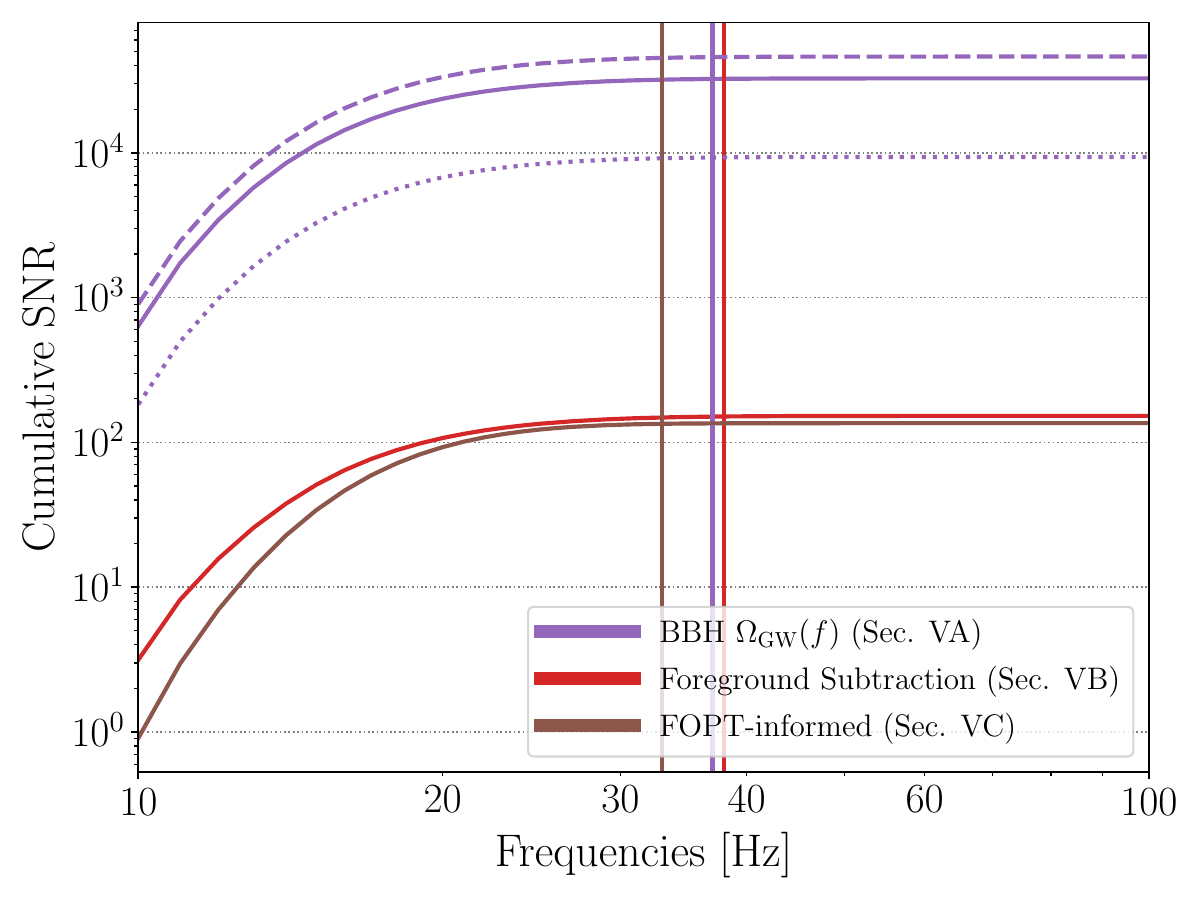}\\
     \caption{ \label{fig:cumulative_SNR} 
     Cumulative signal-to-noise ratio (SNR) curves for the three different injections considered in this work: BBH background (purple), subtraction background (red), and first-order phase transition (FOPT) signal (brown). For reference, we include the scaling of the cumulative SNR on the observation time, for the BBH signal case: 1 month (dotted line), 1 year (full line), and 2 years (dashed line). This dependence scales identically for the different signals. The SNR for stochastic signals employed here is defined as the optimal filter SNR in~\cite{Allen:1999}. Vertical lines mark the frequency values that correspond to where 99\% of the SNR is accumulated.
     }
\end{figure}

Before performing the injection study, we present the cumulative signal-to-noise ratio (SNR) curves for the three different injections considered in this work in Fig.~\ref{fig:cumulative_SNR}, where the SNR for stochastic signals is defined as the optimal filter SNR of~\cite{Allen:1999}. Vertical lines mark the frequency values that correspond to where 99\% of the SNR is accumulated. As may be observed, in all cases frequencies above 40~Hz contribute less than 1\% of the SNR.

For each injection, we simulate $\Chat$ directly, by drawing $\Chat\sim \mathcal N[\Omega_{\mathcal M}(f), \sigma(f)]$, where $\Omega_{\mathcal M}(f)$ is the injected model, and $\sigma(f)$ is the detector noise calculated using the specified noise curves and observation times. For each recovery, we perform $500\,000$ Markov chain Monte Carlo steps, and discard the first $300\,000$ as burn-in. We perform all interpolation in log-space; i.e. we take $x\rightarrow \ln x$, $a \rightarrow \ln a$, etc., when implementing the procedures outlined in Secs.~\ref{sec:intrp_models} and~\ref{sec:proposals}.

\subsection{Full BBH injection} \label{sec:CBC}
We use a BBH $\Omega_{\rm GW}(f)$ as our first example of injection and recover it using this transdimensional spline interpolation approach. %
The contribution to the GWB from all BBHs can be written as
\begin{align}
\label{eq:omgw_astrophys}
    \Omegagw (f) = \frac{f}{\rho_c}\int_{0}^{z_{\textrm{max}}} \textrm{d}z\;\frac{\mathcal{R}(z) \left< \frac{\textrm{d}E}{\textrm{d}f_s}\left.\right|_{f(1+z)}\right>}{(1+z) H(z)}\,,
\end{align}
where $H(z)$ is the Hubble rate, $f_s$ is the frequency of the GW at the source, and $\mathcal R(z)$ is the merger rate of BBHs as a function of redshift. We have used \cite{Callister:2020arv}
\begin{align} \label{eq:source_frame_energy_spectrum}
    \bigg\langle\frac{dE}{df_{s}}|_{f(1+z)}\bigg\rangle = \int dm_1 dm_2 \frac{dE}{df_{s}}&\big(m_1,m_2;f(1+z)\big) \nonumber \\ 
    &\times p(m_1,m_2)\, ,
\end{align}
where we integrate the possible combinations of masses $m_1$ and $m_2$ for a single binary. 
A significant amount of work is done to use individual events~\cite{KAGRA:2021duu}, or joint measurements of individual events and the GWB~\cite{Callister:2020arv, KAGRA:2021kbb, Turbang:2023tjk, Lalleman:2025xcs} to constrain $p(m_1, m_2)$ and $\mathcal R(z)$, where $p(m_1, m_2)$ is the probability distribution of the source masses.
As a note, this curve is generated using \texttt{POPSTOCK} and intended to be an inflated BBH signal curve to emulate the magnitude of the O3 signal \cite{Renzini:2024pxt}.  

The energy spectrum of the source frame radiated by a single binary given in Eq.~\eqref{eq:source_frame_energy_spectrum} is informed by a phenomenological mass distribution model \cite{Abbott_2023}. 
We implement the power-law plus peak (PLPP) model from \cite{Talbot:2018cva}, which has been widely adopted as an astrophysically motivated mass model. 
This model is a truncated PL based on the stellar mass initial function with a Gaussian peak located at a mass $m_{\rm pp}$ that is motivated by a possible overdensity of black holes around 40$M_\odot$. %
A further discussion of the PLPP parameters is provided in \cite{Talbot:2018cva, Renzini:2024pxt}. 

The merger rate as a function of redshift is often assumed to have a similar functional form to the star formation rate (SFR); the canonical argument being that the abundance of black hole binaries, and their subsequent mergers, should track the star formation rate with a time delay between formation and merger, subject to constraints due to the metallicity in the local environment which affects the birth rate of black holes~\cite{Regimbau:2011rp,Rosado:2011kv,LIGOScientific:2016fpe,Callister:2016ewt}. A common choice for this functional form is a smooth BPL similar to the Madau-Dickinson star formation rate, given by~\cite{Callister:2020arv}
\begin{align} \label{eq:R_z_MD}
\mathcal R(z) = \mathcal C(\alpha,\beta ,z_p)\frac{R_0(1+z)^{\alpha}}{1 + \left(\frac{1+z}{1+z_p}\right)^{\alpha+\beta}}\,,
\end{align}
where $R_0$ is the local merger rate, $z_{p}$ is the redshift at which the merger rate peaks, $\alpha$ is the spectral index for $z\lesssim z_{p}$, $\beta$ is the negative of the spectral index for $z\gtrsim z_{p}$, and $\mathcal C$ is a normalization factor to ensure $\mathcal R(0) = R_0$.  For our BBH-like injection, we assume $R_0 = 31.88\, {\rm Gpc}^{-3} {\rm yr}^{-1}$, $ \alpha = 1.9$, $\beta = 3.4$, and $z_p = 2.4$ \cite{Callister:2020arv}.

The argument in favor of using a Madau-Dickinson-like merger rate broadly assumes that the bulk of mergers are from binaries formed in the field. This is in contrast to models like those formed dynamically in dense stellar clusters~\cite{Mapelli_2021, Mandel_2022}, or other hypotheses such as lensed high-redshift stellar-mass BBHs of $\mathcal M_c \lesssim 15\, M_\odot$ where $\mathcal M_c$ is the chirp mass of the binary~\cite{Dominik_2013}. 

The injection is made as if signals, including detectable signals, have been left in the data stream. 
In the next section, we consider the case where detectable signals have been subtracted. 

As stated previously, since the CE sensitivity injections are in the strong signal regime, Eq.~\eqref{eq:variance_on_omega_estimator} cannot be calculated by simply taking the PSD of each detector, as is often done currently. For this work, we assume that we have an independent measurement of $\sigma^2(f)$ (e.g. from a null stream like one might construct from a triangular detector like ET or LISA). However, a future extension of this work 
might be to fit $\sigma^2(f)$ (or the detector PSDs) with our flexible model simultaneously with the background, alleviating the need for an independent measurement.

\begin{figure*}[]
    \includegraphics[width=0.49\textwidth,clip=true]{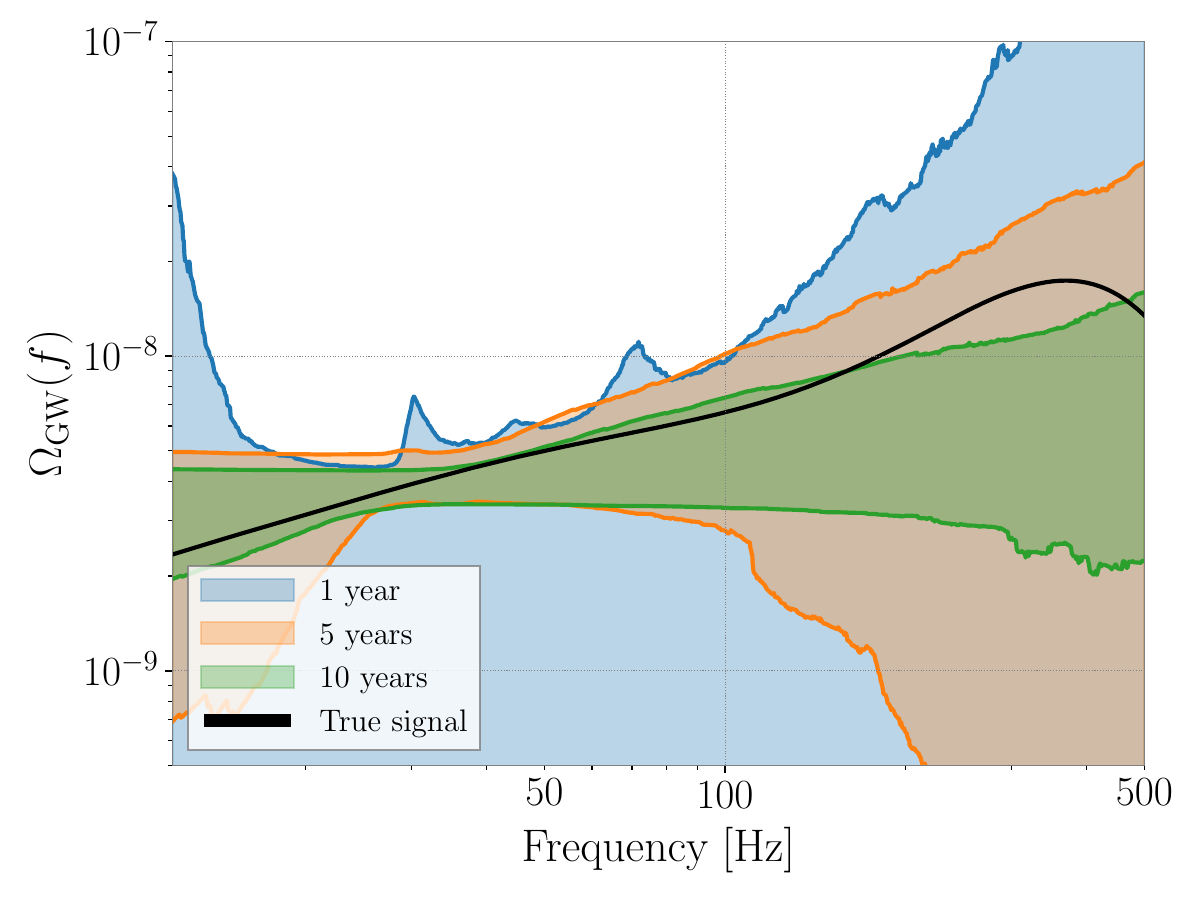}
    \includegraphics[width=0.49\textwidth,clip=true]{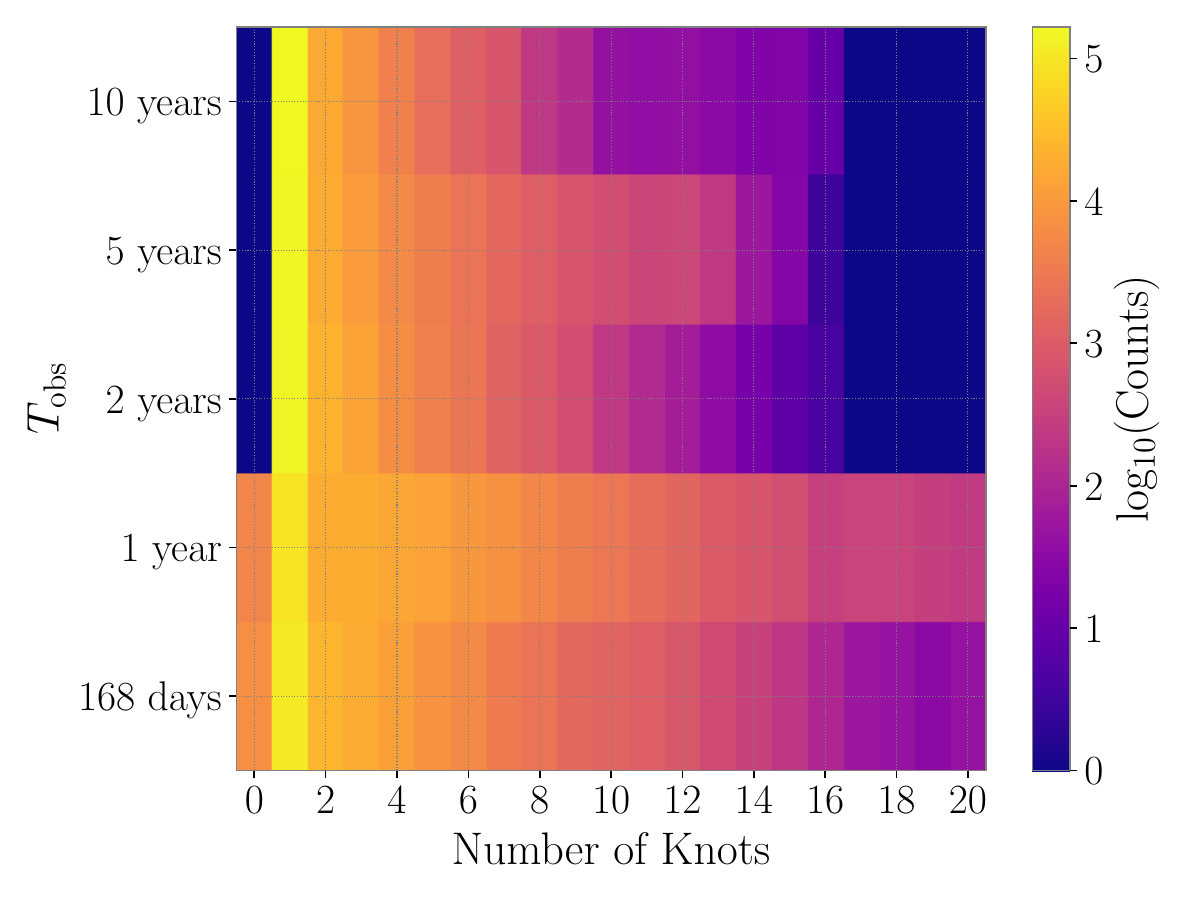}\\
    \caption{ \label{fig:cbc_injection}
    Results for the injection and recovery of a BBH-informed $\Omega_{\rm GW}$, given by Eq.~\eqref{eq:omgw_astrophys}, assuming A+ sensitivity with a range of observation times. 
    On the left, we show the 95\% confidence recovery envelopes for the posteriors of the injected signal (solid black line). 
    On the right, we show the number of knots in each posterior draw for each observation time as a heat map. 
    We find that at two years of observing time, we can confirm a detection of the injected $\Omega_{\rm GW}$.
    However, the range in number of knots marginally decreases as the observation time increases, which signifies that the A+ sensitivity remains low enough for the posterior distributions to fit for noise. }
\end{figure*}

\begin{figure*}[]
    \includegraphics[width=0.49\textwidth,clip=true]{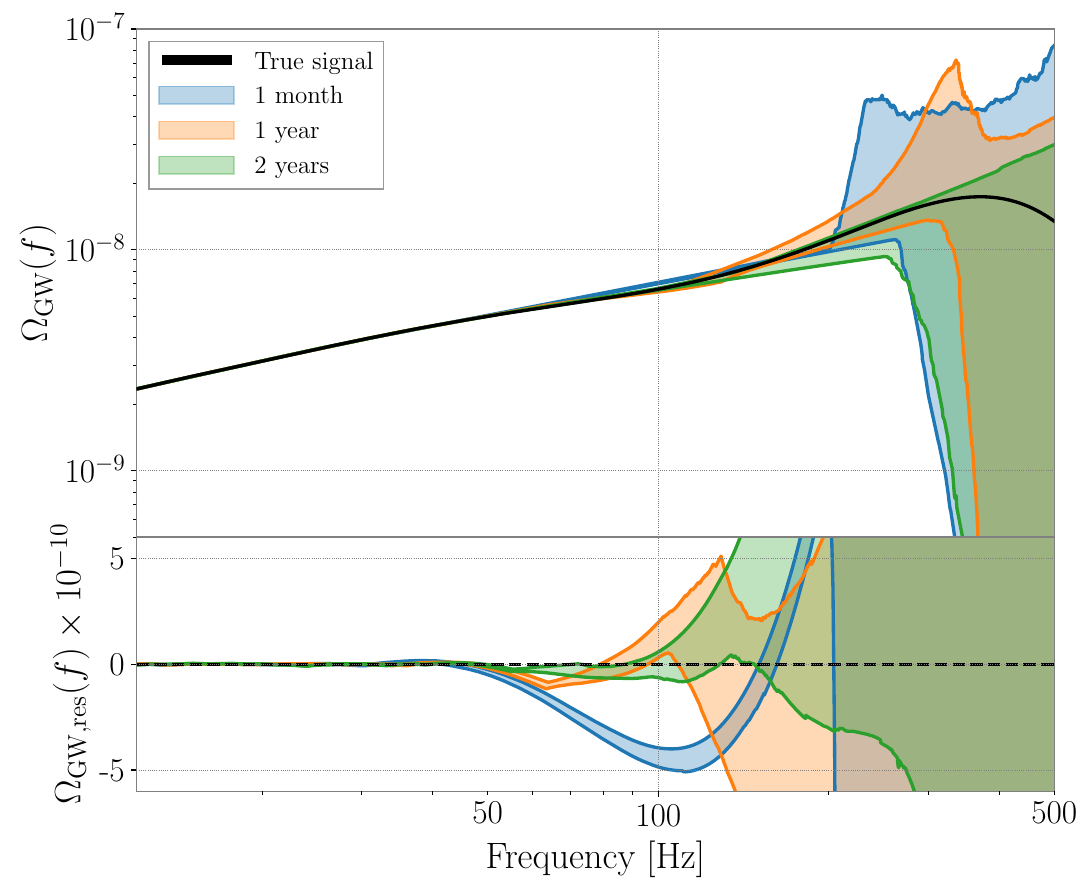}
    \includegraphics[width=0.49\textwidth,clip=true]{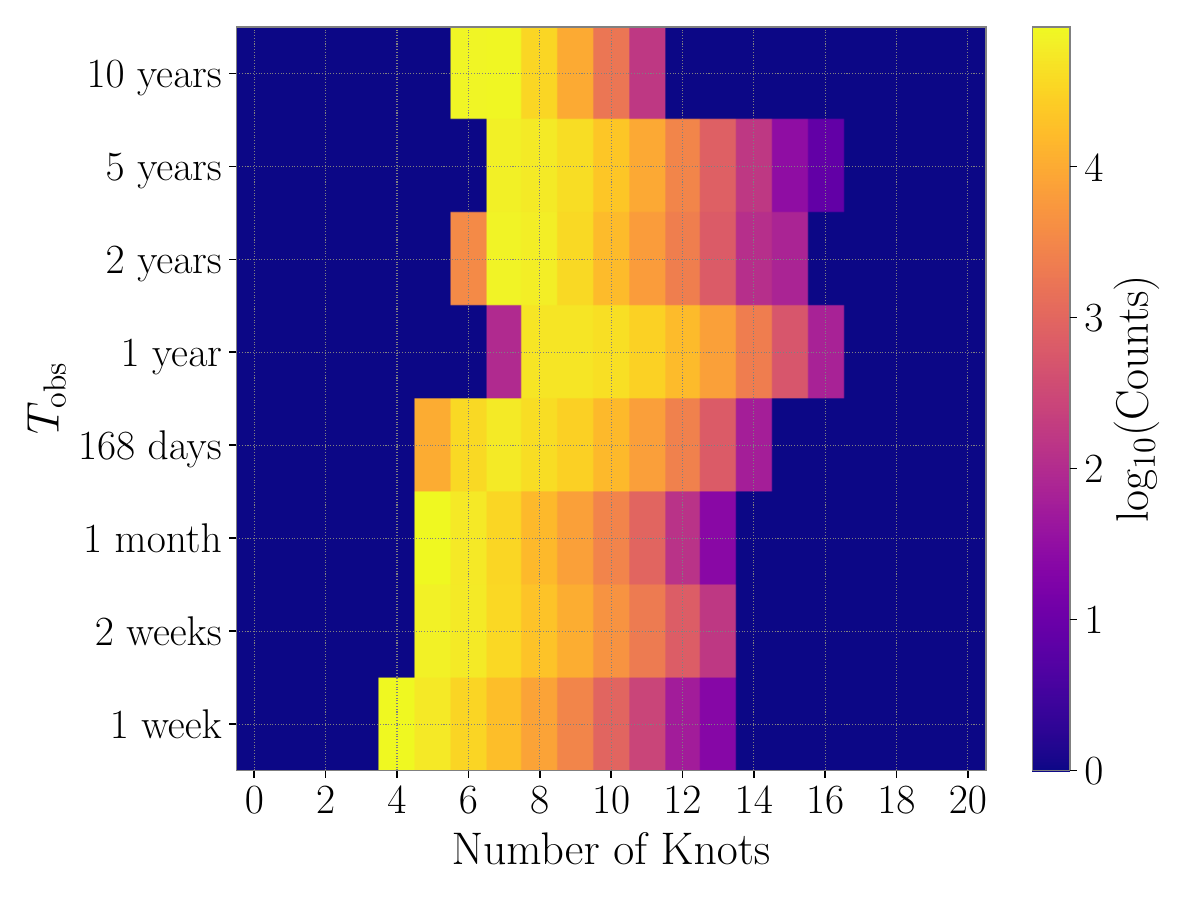}\\
    \caption{ \label{fig:cbc_injection_CE}
    The same injection and recovery as Fig. \ref{fig:cbc_injection} but with a CE sensitivity curve. 
    On the top left, we show the 95\% confidence recovery envelopes for the posteriors of the injected signal (solid black line). 
    The bottom left panel shows the residuals, $\Omega_{\rm GW, res}$, which are obtained by subtracting the $95$\% confidence intervals in the top left panel from the injected signal. 
    On the right, we show the number of knots in each posterior draw for each observation time as a heat map. 
    We are able to recover the injection well up to $f\sim 100$ Hz for all observing times with the number of knots converging as observation time increases.
    }
\end{figure*}

The results of injecting and recovering this BBH background spectrum, $\Omegagw(f)$, with A+ sensitivity are shown in Fig. \ref{fig:cbc_injection} and CE sensitivity in Fig. \ref{fig:cbc_injection_CE}.
We show the 95\% confidence posterior recovery envelopes for various observation times (left) and the corresponding number of knots per posterior (right).
Although we present injections and recoveries for both A+ and CE sensitivity in this section, we only use CE sensitivity in subsequent sections. 

For A+ sensitivity, as shown in Fig. \ref{fig:cbc_injection}, each posterior has an average of one knot until 20 years of observation time, when the average increases to two knots. 
Signal detection is confirmed in less than 20 years with one knot, with a PL trend recovered after 20 years with two knots. 
Even with one knot for shorter observation times, the posteriors still reflect the injection's PL behavior.

Since the number of knots in each posterior sample reflects whether a signal is present within a given sensitivity regime, we can use this information to construct a detection statistic. For example, the right-hand side of Fig.~\ref{fig:cbc_injection} shows that for $T_{\textrm{obs}}>1\,\textrm{yr}$ there are few if any posterior samples with $n=0$, which could indicate a potential detection.
However, simply computing an \textit{ odds ratio}, the number of posterior samples with $n>0$ knots divided by those with $n=0$ knots, is not necessarily sufficient on its own. 
The prior odds built into the odds ratio inherently favor the presence of a signal over noise. 
A potentially more robust detection statistic might undo the inherent prior odds by dividing by the maximum number of knots $N_{\rm max}$, which produces the Bayes factor. However, the extra prior volume of including more available knots (higher $N_{\textrm{max}}$) can result in a Bayes factor that inherently favors noise compared to signal. We discuss these trade-offs in more detail, along with the odds ratio and Bayes factor computations in the context of this transdimensional spline model, in Appendix \ref{apx:bayes}.

The convergence of the number of knots in the posteriors is reflected by the 95\% confidence envelopes narrowing around the injection as we increase the observation time.
For example, the envelopes for 1 and 5 years (green and orange, respectively) of observation time recover the injection for $f \lesssim 100$ Hz. 
Although we can fit the PL behavior of the injection for $f \lesssim 100$ Hz for a wide range of observation times, there is poor recovery at higher frequencies, where A+ detectors are less sensitive to the background signal. 

Similarly to Fig. \ref{fig:cbc_injection} for A+ sensitivity, we present the injection and recovery of a BBH-induced $\Omegagw$ for various observation times with CE sensitivity in Fig. \ref{fig:cbc_injection_CE}. 
Because CE has a greater sensitivity than A+, detection is confirmed earlier for CE sensitivity than for A+. 
The 95\% confidence envelopes recover the $\Omegagw$ behavior almost precisely for observation times as short as 1 month.
Longer observation times improve high-frequency signal resolution, reducing the upper bound of the 95\% confidence intervals to more accurately depict the injection shape, which we also observed for the A+ sensitivity. 

 In Figs.~\ref{fig:cbc_injection} and \ref{fig:cbc_injection_CE}, we see that in regions of the spectrum where the data are uninformative, the model is extrapolated from the parts of the spectrum where is more constrained. In Fig.~\ref{fig:cbc_injection}, this results in a mismatch around the merger-induced peak in the spectrum for the 10 yr case. For Fig.~\ref{fig:cbc_injection_CE}, for 1 month we are unable to capture the late-inspiral-induced ``dip'' in the spectrum around $50$\,Hz because 99\% of the SNR has been accumulated by $\sim 35$ Hz (cf. Fig.~\ref{fig:cumulative_SNR}). Instead, the power law constrained at low frequencies is extrapolated to higher frequencies.

When comparing knot numbers for posteriors, the knots for CE sensitivity posteriors are better constrained per observation time than for A+ sensitivity.
For CE sensitivity, on average $4$-$13$ knots are required to recover the injection, while A+ sensitivity requires anywhere from $1$-$15$ knots. 
Furthermore, as the observation time for CE sensitivity increases to 10 years, the range of knots decreases to $6$-$11$ knots, while the number of knots required for recovering injection with A+ sensitivity with 10 years of observation time is still $1$-$15$ knots. 
Better recovery at higher frequencies, implicated by fewer knots for injections with CE sensitivity, corresponds to better sensitivity to late-inspiral corrections and the $\Omega_{\rm GW}(f)$ turnover. 
We conclude that whether we use A+ or CE sensitivity affects the convergence of the number of knots in each posterior, in addition to the accuracy of the recovery envelopes.   

This example of BBH GWB injection and recovery illustrates the connection between observation time and the effectiveness of the posterior recovery of the injection. 
We find that our transdimensional spline interpolation model can identify a detection and resolve features as the observation time increases. 

\subsection{Foreground event subtraction} \label{sec:subtraction}

In this section, we consider the situation where resolvable foreground CBC events have been subtracted, and the unresolved GW background after this subtraction is the dominant contribution to the measured spectrum.
Subtraction is typically performed by removing contributions from resolvable CBCs~\cite{Sachdev:2020bkk, Regimbau:2017, Harms_2008, Crowder_2005, Cutler:2005qq,Zhong:2022ylh,Zhong:2024dss,Zhong:2025qno}, and we consider the scenario in ~\cite{Sachdev:2020bkk} where there is also some contribution to the signal from subtraction error.
With next-generation detectors, it is expected that a fraction of BNSs and all but the furthest BBHs are resolvable, so the expected dominant contribution to the unresolved background, will likely be the residual BNS signal.  

For the reasons discussed above, we choose to inject the $\Omega_{\rm BNS, residual}$ curve from Fig. 1(e) of \cite{Sachdev:2020bkk}. This is the combined signal from the unresolved BNS background and foreground subtraction error in the HLV 3G three-detector network. The curve corresponds to subtraction of foreground events that have a network SNR greater than 8. This signal is interesting because it has a high-frequency turnover similar to the CBC injection from the previous section, in addition to a low-frequency feature due to the limited detector sensitivity at those frequencies. 

\begin{figure*}[]
    \includegraphics[width=\textwidth,clip=true]{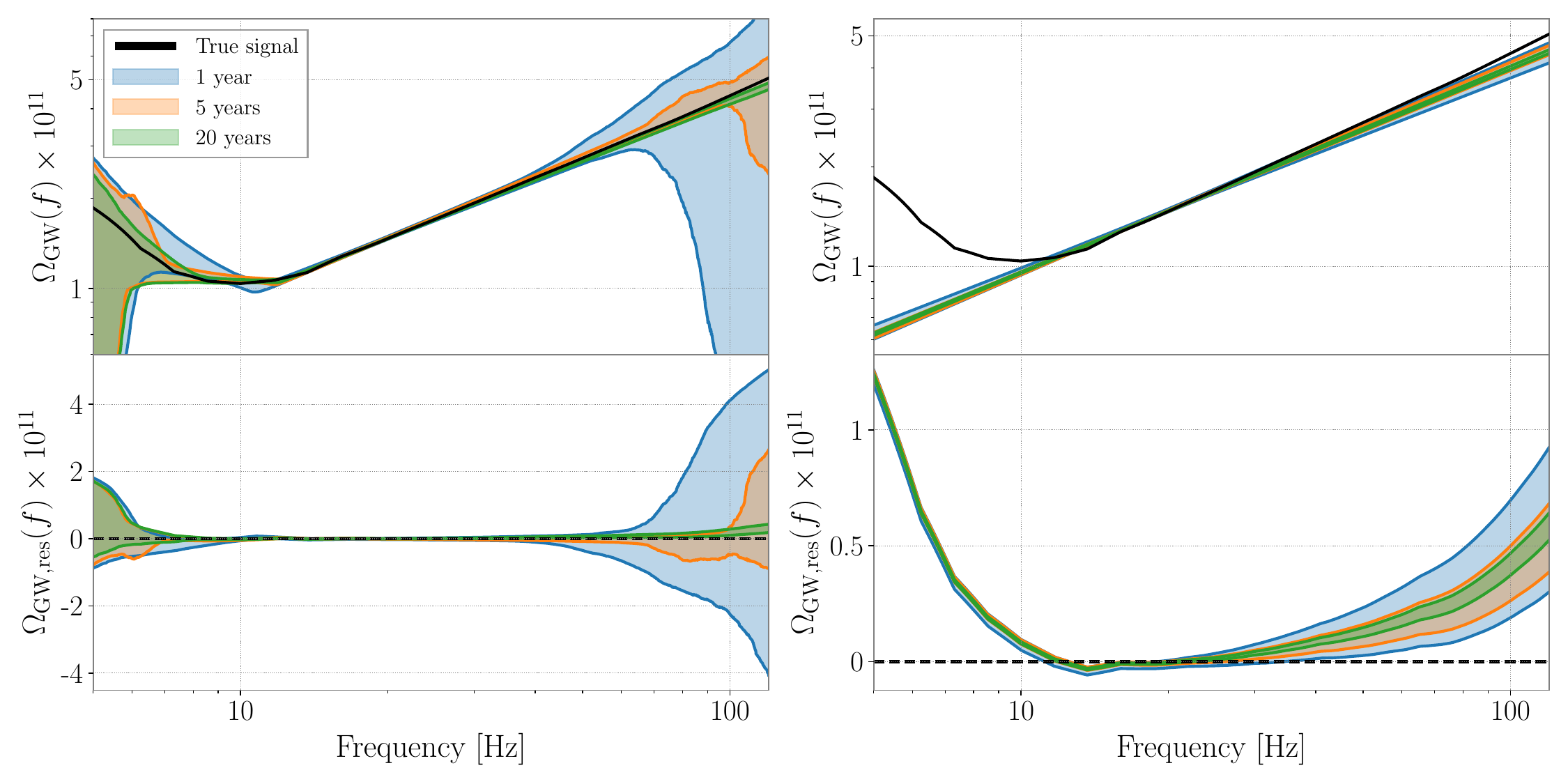}\\
    \caption{ \label{fig:sachdev_1e} 
    Recovered posteriors of the unresolved BNS signal $\Omega_{\rm BNS, residual}$ after subtracting resolved signals with SNR$>8$ using CE noise. The injected signal we use is from~\cite{Sachdev:2020bkk}. 
    The left plots shows the recovered posteriors for $N_{\rm max} = 30$ and the top right plots shows the recovered posteriors for $N_{\rm max} = 2$. 
    Each color envelope represents 95\% confidence for the recovered posteriors for various observation times for the simulated CE noise (legend).
    The bottom plots provide the residuals $\Omega_{\rm BNS, residual;res}$ for the 95\% confidence envelopes subtracted from the injected signal,  $\Omega_{\rm BNS, residual}$, from the top row. 
    Limiting $N_{\rm max} = 2$ reveals PL posteriors that track the injection at later frequencies, while allowing $N_{\rm max} = 30$ detects a low-frequency subtraction feature. 
    Mostly positive residuals $\Omega_{\rm BNS, residual;res}$ indicate that fitting a non-PL signal with a PL biases the fit, causing posterior draws miss sections of the injection and consequentially underfit the PL portion.
    }
\end{figure*}

The population of BNSs used in \cite{Sachdev:2020bkk} is generated using a Monte Carlo procedure and fiducial parameter distributions \cite{Regimbau:2016ike, Regimbau:2011rp, Meacher_2015, abbottGW170817ImplicationsStochastic2018}. 
The full $\Omega_{\rm BNS}$ signal is the superposition of 1,438,835 BNSs for one year of observation using an SFR-informed redshift distribution with local BNS merger rate $R(z=0) = 920\, {\rm Gpc}^{-3} {\rm yr}^{-1}$ \cite{Abbott_2019}, intrinsic source frame masses drawn from $\mathcal N(1.33\,M_\odot, 0.09\, M_\odot)$, and BNS sky locations and angular parameters drawn from uniform distributions. 
We refer the reader to \cite{Sachdev:2020bkk} for more details on the BNS population used to generate the full $\Omega_{\rm BNS}$ signal that we obtain $\Omega_{\rm BNS, residual}$. 
Although more detailed Monte Carlo analyses that quantify the strength of the unresolved background using more recent population models and rates have been performed since those in \cite{Sachdev:2020bkk} (see, e.g., \cite{Bellie:2023jlq}), we proceed with $\Omega_{\rm BNS, residual}$ for its notable low-frequency feature relevant to this study.

We inject the signal $\Omega_{\rm BNS, residual}$ with a CE sensitivity curve, assuming $T_{\rm obs}=1,\,5,\,20\,\textrm{yr}$, and present the results of these recoveries in Fig. \ref{fig:sachdev_1e}. 
To start, we allow $N = 30$, which provides the interpolation model with the most flexibility. We maintain linear interpolation in log space between the recovered knots. The results of the 95\% confidence envelope for a range of $T_{\rm obs}$ (legend) are shown in the left panel of Fig. \ref{fig:sachdev_1e}. 
For each of the observation times, the posteriors (top left) recover the PL behavior from $\sim 15$-$60$ Hz and detect the upturn at $\sim 10$ Hz. 
As the observation time increases, the lower bounds of the 95\% confidence envelopes align more closely with the injection, while the upper bounds effectively detect the injected signal for each observation time.
In general, increasing the maximum number of knots allows the posteriors to recover more features with the trade-off of overfitting regimes of less sensitivity. For example, at high frequencies, where we have less sensitivity, the 95\% confidence envelopes are wider. 

We now choose $N_{\rm max} = 2$, probing the effects of fitting a non-PL signal with a PL signal using our model. 
To evaluate the effects of applying a PL to $\Omega_{\rm BNS, residual}$, we implement linear interpolation in the posteriors, which are shown for the same observation times as for $N_{\rm max} = 30$ (top right) along with the residuals between the injected signal and the posteriors, $\Omega_{\rm BNS, residual;res}$ (bottom right).
The PLs recovered in the top right panel of Fig. \ref{fig:sachdev_1e} are consistently shallower than the PL portion of the injection. 
The residuals between the signal and the posteriors confirm this, as the 95\% confidence envelopes on the residuals are mostly positive. 
Therefore, by trying to fit an incorrect or overly constrained model to an injection using this transdimensional spline interpolation model, we will obtain posteriors with a bias toward the ignored features, leading to a recovery that is not truly accurate anywhere. 
In fact, fixing a PL model as sensitivity increases may risk recovery bias with any potential recovery algorithm, not just with this spline interpolation model.
So, implementing this spline interpolation model with its full flexibility is the best way to consistently and correctly recover an injected signal.

\subsection{First-order phase transition-informed BPL} \label{sec:fopt}

\begin{figure*}[]
    \includegraphics[width=0.49\textwidth,clip=true]{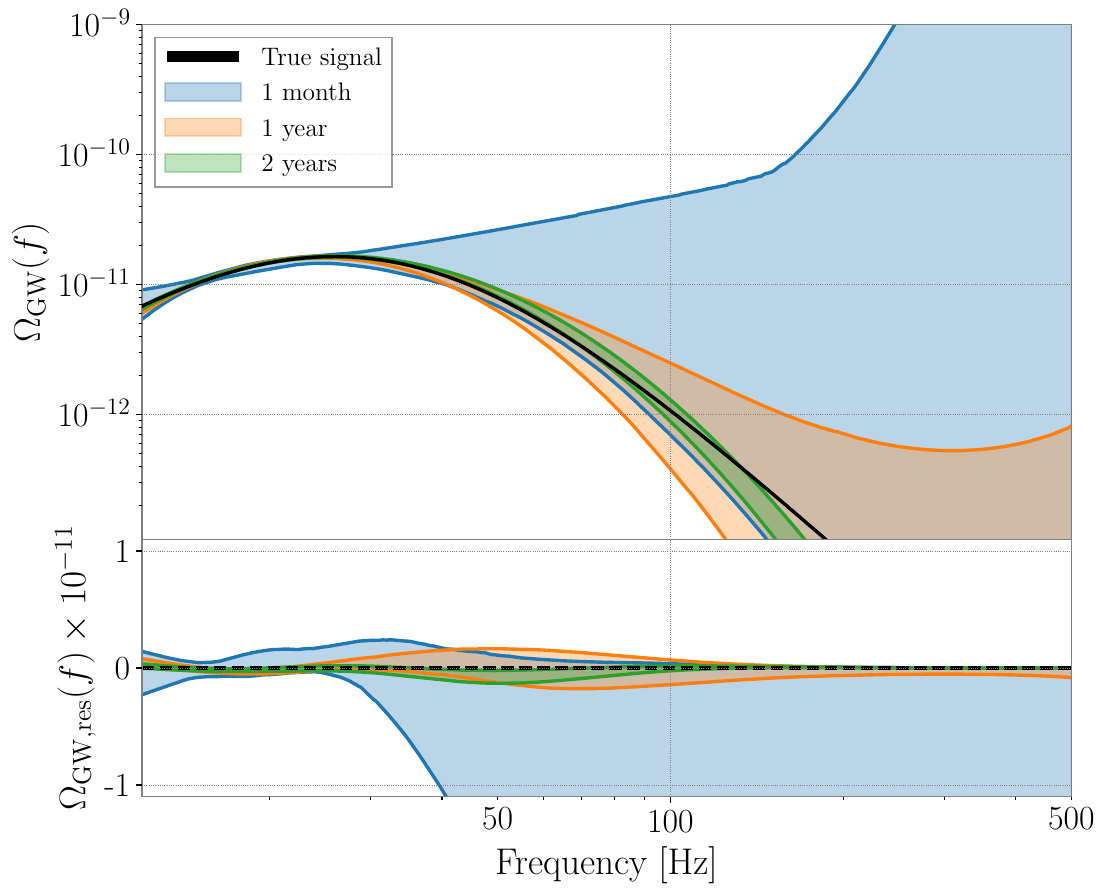}
    \includegraphics[width=0.49\textwidth,clip=true]{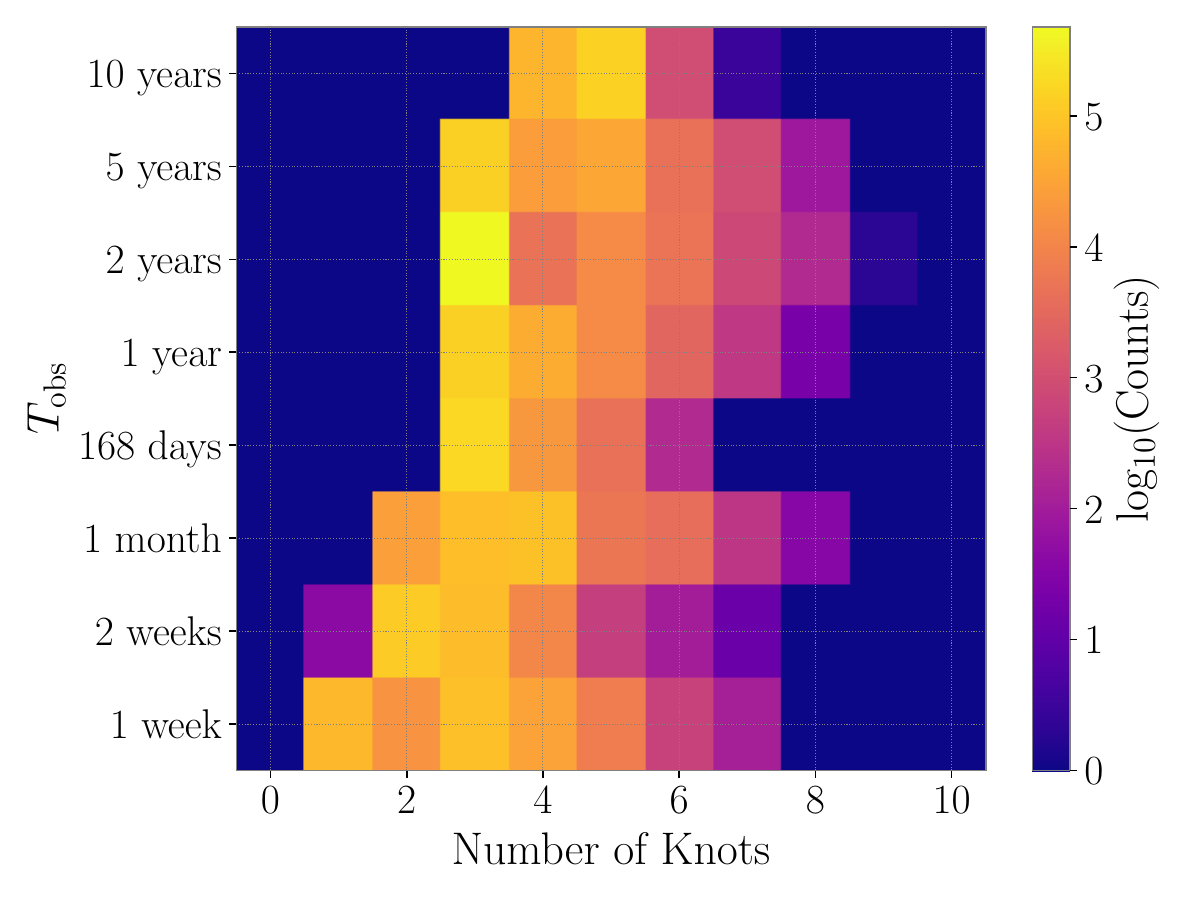}\\
    \caption{ \label{fig:posteriors_FOPT_CE} 
    Results for injecting and recovering a FOPT-informed smoothed BPL with a CE sensitivity curve. The injected signal is described in Eq.~\eqref{eq:FOPT} from \cite{Sachdev:2020bkk}.
    We show the 95\% confidence envelopes for the recovered posteriors for observation times ranging from 1 week to 2 years on the top left. The injected signal is overlaid (black solid line). 
    The bottom left panel shows the residuals, $\Omega_{\rm GW, res}$, which are obtained by subtracting the $95$\% confidence intervals in the top left panel from the injected signal. 
    On the right, we show a heat map of the number of knots in each posterior (color bar) as a function of observation time. 
    As observation time increases, we can better constrain the recovered posteriors. 
    }
\end{figure*}

\begin{figure}[]
    \includegraphics[width=\columnwidth,clip=true]{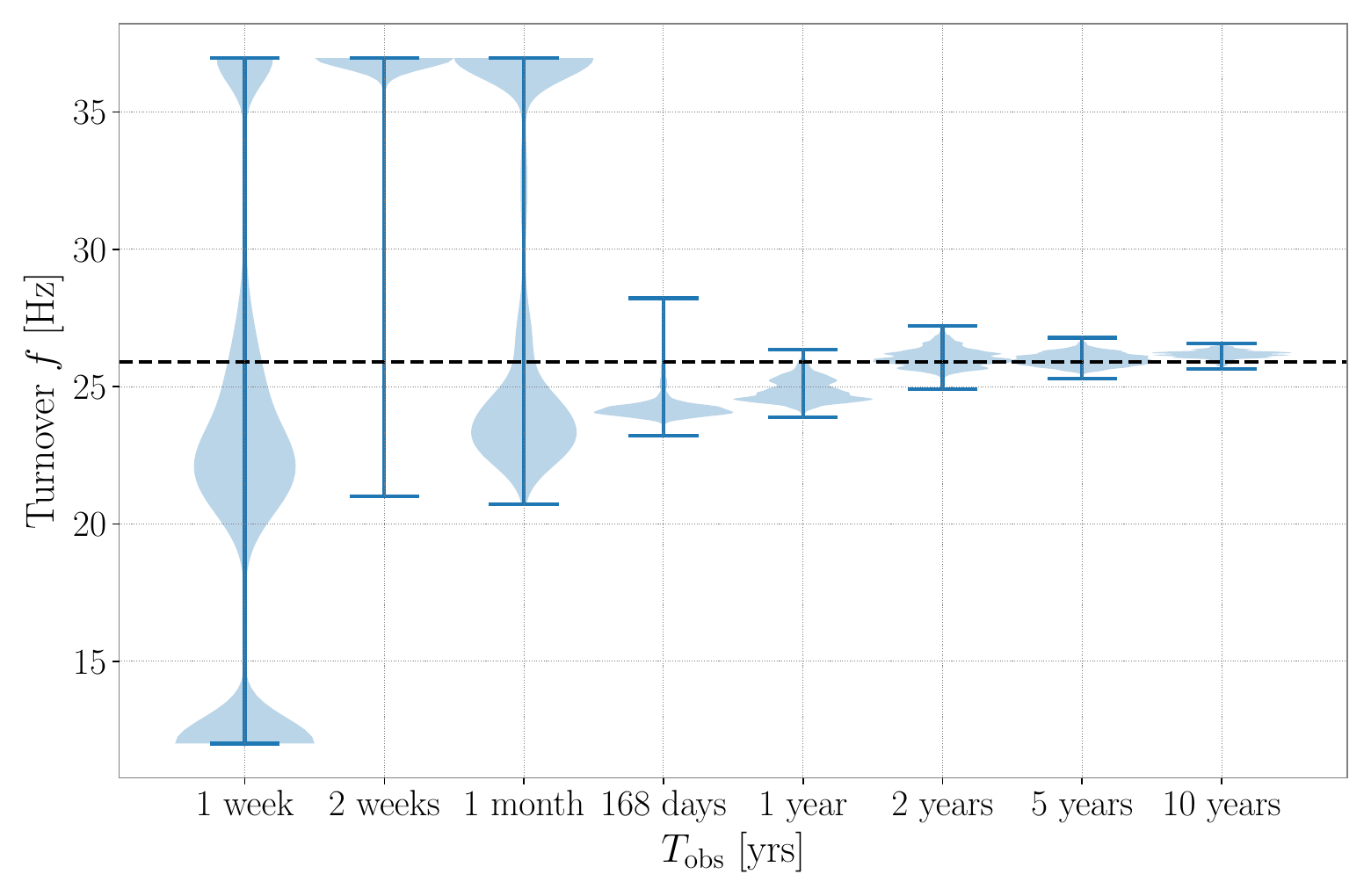}\\
    \caption{ \label{fig:turnover_FOPT_CE} 
    Violin plots showing the recovered turnover frequency as a function of observation time for injected and recovered FOPT-informed smoothed BPL with CE sensitivity. 
    These turnover frequencies are computed from the same posteriors whose results are summarized in Fig. \ref{fig:posteriors_FOPT_CE}.
    For each posterior and observing time, we compute the frequency where $\Omega_{\rm GW}$ obtains its maximum, deeming that the turnover frequency. 
    For observation times at a month and shorter, the violin plots are multimodal with peaks around $36$ and $13$ Hz, which correspond to posteriors that do not turn over. 
    As observation time increases, the posterior turnover frequencies converge at the injected frequency (black dashed line). 
    }
\end{figure}

We next inject a smoothed BPL, a common signal shape that arises when predicting the shape of a primordial GWB spectrum~\cite{Caprini:2007xq,Huber:2008hg,Martinovic:2020hru,Romero:2021kby}. Specifically, a BPL is a reasonable fit to the predicted spectrum from first-order phase transitions (FOPT) in the early Universe. It is also a crude model for the stellar-mass CBC background, which is a PL that turns over at high frequencies. 

FOPTs in the early universe involve the formation of lower-temperature bubbles of different phases of matter or fields that expand into their surrounding medium. 
FOPTs produce GWs through three main mechanisms: (i) bubble collisions that generate shock waves that produce GWs \cite{Hindmarsh:2013xza}; (ii) sound waves in the plasma from bubble expansion, which are probably the dominant FOPT GW source \cite{Weir_2018}; and (iii) plasma turbulence induced by bubble motion, especially during strongly first-order electroweak transitions \cite{Huber:2008hg}. 

There are many analytical and numerical models for each FOPT effect that produces GWs. As in \cite{Martinovic:2020hru}, we choose to model the frequency spectrum as a smoothed BPL:
\begin{equation} \label{eq:FOPT}
   \Omega_{\rm FOPT} = \Omega_* \left( \frac{f}{f_*} \right)^{\alpha_1} \left[ 1 + \left( \frac{f}{f_*} \right)^\Delta \right]^{(\alpha_2 - \alpha_1)/\Delta}\,.
\end{equation}
As in \cite{Martinovic:2020hru}, we use parameters $\alpha_1 = 3,\,\alpha_2 = -4,\, \Delta = 2,\, f_* = 30,$ and $\Omega_* =1.8\times 10^{-10}$.

We inject the model from Eq.~\eqref{eq:FOPT} with the CE sensitivity curve and a range of $T_{\rm obs}$ from 1 week to 10 years. Because we expect a smooth turnover, we interpolate using cubic splines. We present and compare the results of using linear and Akima spline interpolation in Appendix~\ref{apx:interp}. We allow a maximum of 30 knots for each recovered posterior. 

We show the results of a smooth FOPT-informed BPL injection and recovery in Fig. \ref{fig:posteriors_FOPT_CE}. The leftmost panel shows the recovery envelopes for various $T_{\rm obs}$. As we increase $T_{\rm obs}$, the 95\% recovery envelope of posteriors narrows and better tracks the injected signal (solid black line). By an observation time of 2 years (green), the recovery envelope identifies the full shape of the injection. 

Not only can we use our model to recover an injected signal, but we can also extract features of the recovered spectrum. One such parameter is $f_{\rm turn}$, which is the frequency at which the smoothed BPL is at its maximum. We define $f_{\rm turn}$ as the frequency of the maximum signal value:
\begin{equation}
    f_{\rm turn} = f_* \left| \frac{\alpha_1}{\alpha_2} \right|^{1/\Delta}
\end{equation}
by taking the derivative of Eq.~\eqref{eq:FOPT} with respect to frequency. 

We show the recovery of the BPL turnover frequency $f_{\rm turn}$ in Fig. \ref{fig:turnover_FOPT_CE}. 
Each violin is a histogram of the $f_{\rm turn}$ for 50,000 posterior samples for each $T_{\rm obs}$. 
The ends of the violins are the extrema for each recovery histogram. 
We find that we can recover $f_{\rm turn}$ that peak near the injected turnover (black dashed line) for each observation time.
For one month or less of observation time, the histograms have a peak beyond $35$ Hz, as well as a peak near the injected $f_{\rm turn}$. 
This means that some of the posteriors are PLs or diverge for shorter observation times. 
As the observing time increases, the posteriors peak near the injected $f_{\rm turn}$ and the extrema narrow around the injected value. 

Similarly to Sec. \ref{sec:CBC}, we can examine the distribution of the number of knots used in the recovered posteriors. 
At only 1 week of observing time, posteriors have anywhere from one to four knots, but as we increase the observing time, we obtain posteriors with mostly three knots. 
This shift occurs after 30 days of observation time, which matches the behavior of the posterior envelope in the left panel of Fig. \ref{fig:posteriors_FOPT_CE}. 
We can conclude that the fitter opts for the minimum required number of knots to produce an accurate posterior for the injected signal.

Similarly to the subtraction example in the previous section, we find that we can recover a more complicated injection than a PL. 
We show that even with less observation time than usual, we can still recover key features of injection, for example $f_{\rm turn}$, from the posteriors, while increasing sensitivity and observation time improve the results.

\section{Constraining the merger rates of CBCs}\label{sec:constrain_Rz}
With the increasing number of CBC detections, we can better determine the properties of their underlying populations, and ultimately study the populations of their massive star progenitors~\cite{Zevin:2017evb,Zevin:2020gbd}. 
Past work has illustrated a joint hierarchical analysis involving resolved and unresolved signals to place constraints on the merger rate of black holes as a function of redshift~\cite{Callister:2020arv}. 
Recently, new methods have been proposed to constrain the GWB from black hole binaries in dense stellar clusters~\cite{Kou:2024gvp}, and to constrain the metallicity dependence of black hole formation and the distribution of time delays between black hole formation and merger~\cite{Turbang:2023tjk}.

\begin{figure}[]
    \includegraphics[width=\columnwidth,clip=true]{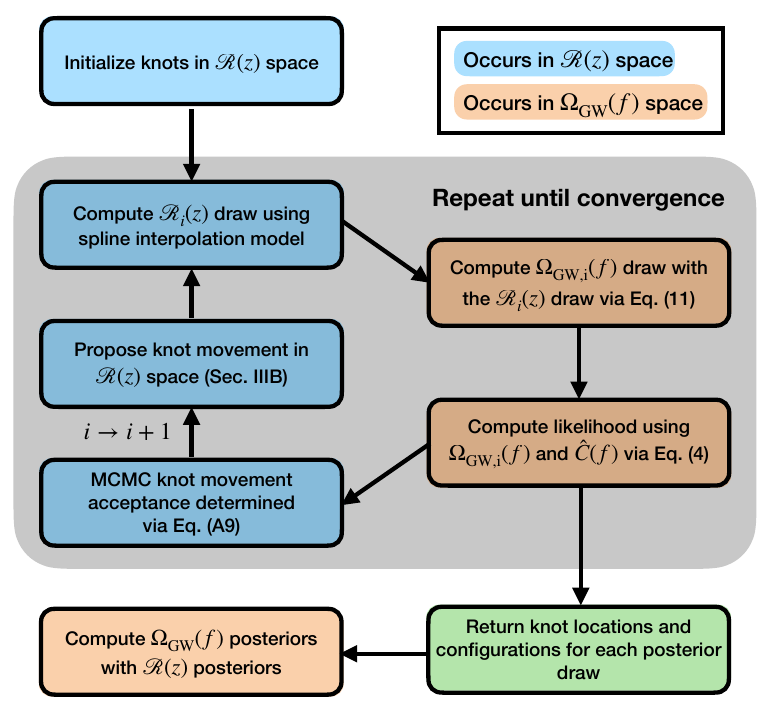}\\
     \caption{ \label{fig:R_z_schematic} 
     A flowchart for the hierarchical implementation of the spline interpolation model in Sec. \ref{sec:constrain_Rz}.
     Blue (orange) boxes indicate actions that occur in the $\mathcal R(z)$ [$\Omega_{\rm GW}(f)$] space, and the green box is where the sampler returns the returns of the RJMCMC sampling of the spline interpolation model.
     Boxes in the gray region are the steps that occur within the hierarchical sampling and model implementation. 
     By moving between the $\mathcal R(z)$ and $\Omega_{\rm GW}(f)$ spaces in the sampler, we can perform hierarchical fitting using our transdimensional spline interpolation model. 
     }
\end{figure}

\begin{figure*}[]
    \includegraphics[width=0.49\textwidth,clip=true]{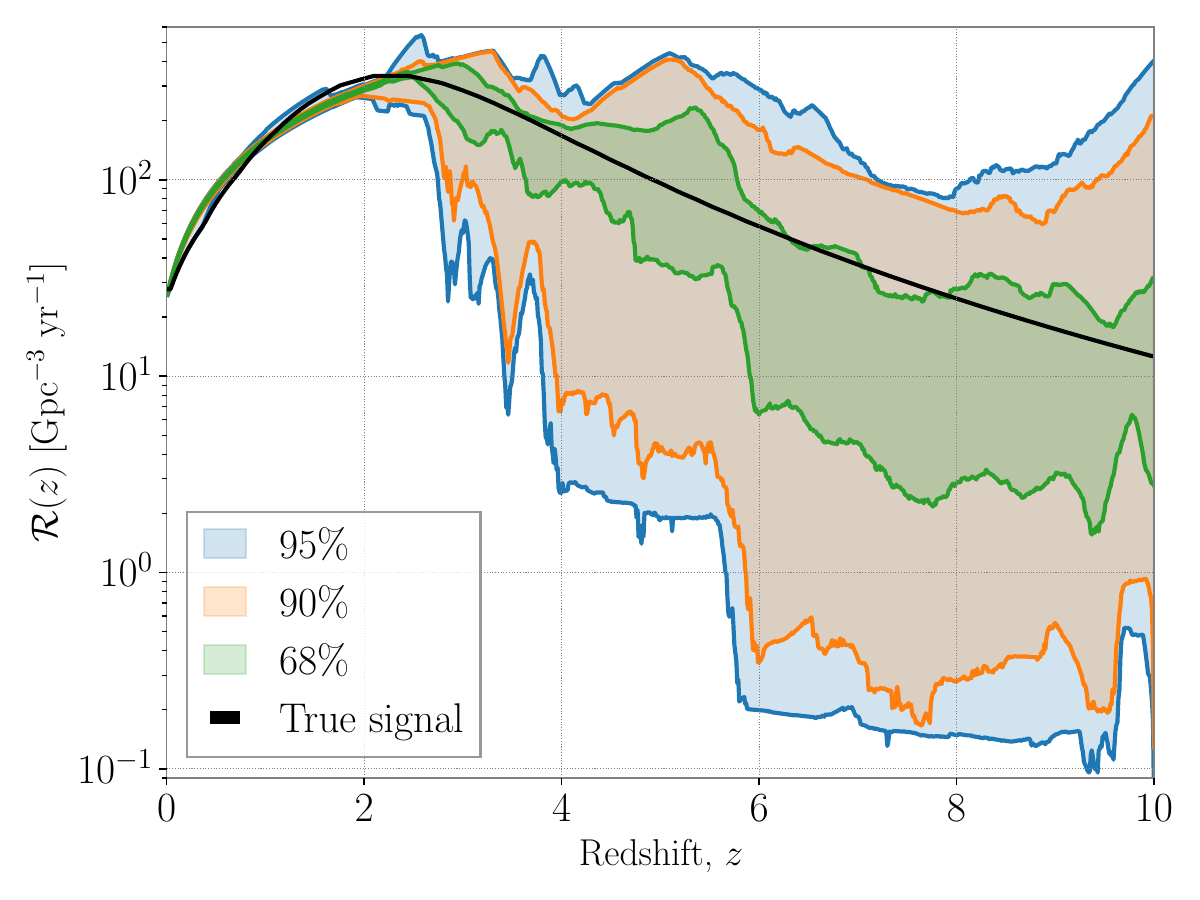} 
    \includegraphics[width=0.49\textwidth,clip=true]{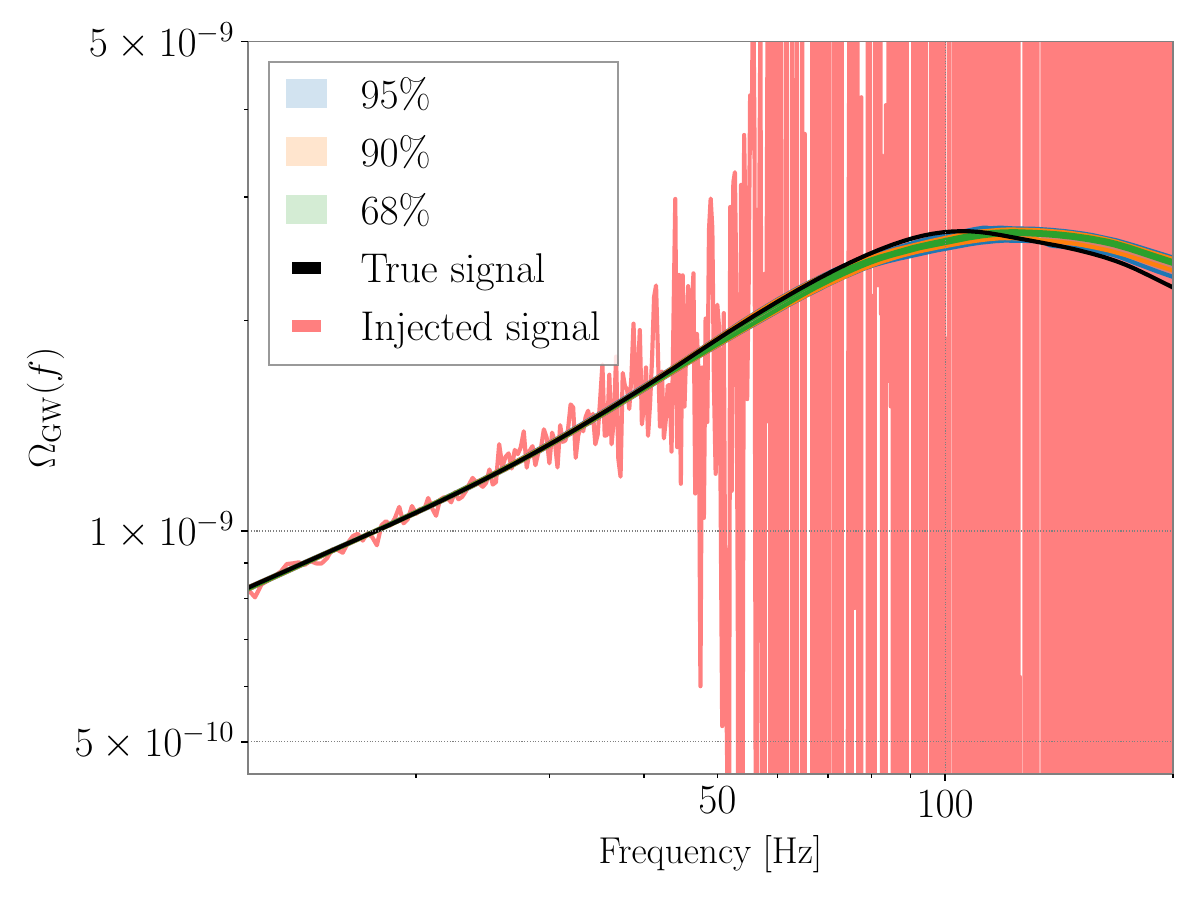}\\
    \caption{ \label{fig:R_z_recovery} Results for injecting an $\Omega_{\rm GW}$ and recovering both the $\Omega_{\rm GW}$ and its hyperparameter $\mathcal R(z)$ from Eq.~\eqref{eq:omgw_astrophys}.
    On the left, we show the recovery envelopes for an injected $\mathcal R(z)$ (black) with 68\%, 95\%, and 99.7\% confidence envelopes (blue) corresponding to 1, 2, and 3 standard deviations, respectively. The right plot shows the 95\% confidence recovery envelope (blue) for the corresponding $\Omega_{\rm GW}$ (black) with CE sensitivity (red). 
    Although the posteriors have deviations from the smooth injected $\mathcal R(z)$, we still recover the correct turnover shape of the injected Madau-Dickinson-informed $\mathcal R(z)$ and the correct $\Omega_{\rm GW}$ using our transdimensional spline interpolation algorithm.
    }
\end{figure*}

Here, we employ this transdimensional algorithm to directly infer a BPL merger rate model, as in Eq.~\eqref{eq:R_z_MD}.
This approach demonstrates that hierarchical analysis can be conducted effectively on a detected background signal with next-generation detectors, thereby revealing further insights that can be extracted from an observed GWB.
For this merger rate mode, we utilize parameters $R_0 = 25\, {\rm Gpc}^{-3} {\rm yr}^{-1}$, $ \alpha = 2.7$, $\beta = 3.4$, and $z_p = 2.4$ \cite{Callister:2020arv}.
In this injection, we implement the same CE sensitivity curve used in Sec. \ref{sec:fit_omgw} with 12 hours of observing time.

We point to Fig. \ref{fig:R_z_schematic} for the work flow of how we inject an $\Omega_{\rm GW}(f)$ signal and utilize our spline interpolation model to recover the true $\mathcal R(z)$ associated with the $\Omega_{\rm GW}(f)$ injection. 
First, similar to the examples in Sec. \ref{sec:fit_omgw}, we place the interpolating knots in redshift space and implement a few constraints.
The knot $z$-value priors are placed linearly in the redshift space with a maximum of 30 knots that range in placement from $z=0$ to $z=10$, and their amplitudes are log-uniformly distributed between 0.1 and $10^3$.
We do not allow knots to be placed at $z>10$ because we assume that no mergers occur earlier than $z=10$ \cite{Callister:2020arv}.
Second, we do not fix the number of knots because we want to test the full flexibility of the model and its ability to recover signals with an efficient number of knots. 
However, we fix the amplitude of the first knot at $z = 0$ because $\mathcal R(z=0)$ is astrophysically well constrained. 
The $\mathcal R(z)$ that we inject is a perfect BPL, and we would like to test whether the model can recognize the BPL and implement an appropriate number of knots. 

Next, we begin the RJMCMC sampling step of the spline interpolation model (gray box in Fig. \ref{fig:R_z_schematic}).
Interpolating the knots placed in the redshift space, we compute $\mathcal{R}_i(z)$. 
This is used to compute a corresponding $\Omega_{\rm GW,i}(f)$ via Eq.~\eqref{eq:omgw_astrophys}. 
This $\Omega_{\rm GW,i}(f)$ is finally used to compute the log-likelihood in Eq.~\eqref{eq:ln_likelihood}.
After the log-likelihood for this proposed point has been computed, we compare it to the log-likelihood of the current point and decide whether to accept the new point. Then, we propose a new knot movement in the redshift space to produce the next proposed point, $\mathcal R_{i+1}(z)$. 
This procedure is repeated until convergence or the specified maximum number of samples is reached. 

The results of this injection and recovery for both $\Omegagw(f)$ and $\mathcal{R}(z)$ are shown in Fig. \ref{fig:R_z_recovery}. 
On the left, we show the redshift space where we have recovered the $\mathcal{R}(z)$. 
On the right, we show the frequency space where we have calculated the corresponding injected $\Omegagw$ (black line) with CE sensitivity (red). 
The 12-hour observing time, even with the CE sensitivity curve, limits the high-frequency sensitivity, causing large amounts of noise in the injected signal after $\sim 50$ Hz.
We simulate 12 hours of observation time because it is the shortest duration that minimizes numerical error from computing $\Omegagw (f)$ with \texttt{POPSTOCK} relative to the CE sensitivity curve.
Recovery envelopes for various confidence intervals are shown in various overlaid colors (legend).

With this hierarchical implementation of our spline interpolation model, we can recover important features of \textit{both} the functional spaces. 
In the redshift space where we place the interpolating knots and employ our model (left), for each recovery envelope, we constrain the PL behavior for  $z \lesssim z_p$. 
For each confidence envelope, especially the $68\%$ confidence (green), we recover the $z \lesssim z_p$ regime well and observe a turnover in the posteriors around $z = z_p$.
After $z=z_p$, the $68\%$ confidence envelope, while not as constrained as at low redshifts, still tracks the injected $\mathcal R(z)$ behavior. 
On the other hand, the $90\%$ (orange) and $95\%$ (blue) recovery envelopes return the prior on the knot amplitudes in redshift space for $z \gtrsim  z_p$, indicating that this redshift regime is poorly constrained. 
Returning the prior at high redshifts also implies that variations in the $\mathcal{R}(z)$ posteriors at higher redshifts produce subtle effects on $\Omegagw(f)$ that we are not sensitive to with just 12 hours of data. 

Because we apply the spline interpolation model in the redshift space and not in the frequency space, the recovered $\Omega_{\rm GW}(f)$ is determined differently than in the previous examples (Sec. \ref{sec:fit_omgw}). 
In previous injection and recovery examples, we place the knots in the frequency space, which extracts the $\Omega_{\rm GW}(f)$ at each frequency independently from the adjacent frequencies. 
For the example in this section, we place the knots in the redshift space, which do not directly map to knots in the frequency space. 
For this reason, changes to the knots, and subsequently the interpolation model, in the redshift space induce changes to $\Omega_{\rm GW}(f)$ at a number of frequencies. 
This effect is not limited to our choice to fit $\mathcal R(z)$ with our spline interpolation model--we expect to see an equivalent effect if we were to instead fit the mass probability distribution [Eq.~\eqref{eq:source_frame_energy_spectrum}].
The successful hierarchical analysis of this example demonstrates the versatility of our model in fitting an $\Omega_{\rm GW}(f)$ signal across different sensitivity regimes and extracting crucial astrophysical information from signals.

\section{Discussion and Conclusions}
\label{sec:disc_and_conc}

Future GW detectors are expected to be sensitive enough to enable the detection of a GWB signal. %
In this work, we have implemented a flexible RJMCMC model to infer the GWB spectrum in a series of scenarios. %
Then, assuming a GWB produced by compact binaries, we employ the same transdimensional interpolating spline model to infer the underlying merger rate of CBCs directly. %
Although the GWB signal can often be described by a PL to a good approximation, there are several cases where we can expect deviations from the canonical PL model. %
The same holds for the Madau-Dickinson model for the merger rate of BBHs. %
Therefore, the development of versatile tools like this one will become increasingly important as detectors become more sensitive.

We perform three different representative injections and show their recoveries in Sec. \ref{sec:fit_omgw}, to showcase the capabilities of our method in the context of GWB searches. 
We first inject a realistic BBH background, consistent with current BBH population models, employing a LIGO A+ noise curve. As a reminder, no individual signals were subtracted or removed. 
Increasing the observing time improves the degree to which we can recover injections, but nonetheless the A+ sensitivity will not be enough to resolve the high-frequency turnover. 
Therefore, we turn to injections using CE sensitivity and find that, for all three example injections, we can consistently recover key features of the injected signals using this model. 

We proceed with the CE sensitivity curve and inject a background that emulates the residual after subtracting resolvable BNSs. With sufficient observation time, we can resolve low-frequency features, such as those introduced by the subtraction method. 

Next, we inject a smoothed BPL, which is a type of signal we expect to see from, e.g., FOPTs in the early Universe.
We are able to accurately recover the injected signal, and also use the flexible model to place posteriors on model parameters. 
In this example, we construct posteriors on the recovered turnover frequency of the injected smoothed BPL. This showcases the power of the flexible algorithm, which could be applied in many other cases. 
Additionally, we find that even if the model were given flexibility to choose the number of interpolating knots, it tends to minimize the necessary number of knots, reducing the prior volume dynamically, and eliminating the need to manually test a number of possible models of increasing complexity. 

Finally, we employ our algorithm to directly infer the merger rate of BBHs as a function of redshift and use this to construct a model for $\Omegagw(f)$. As shown in Sec. \ref{sec:constrain_Rz}, we can recover the injected $\mathcal R(z)$ using our flexible model assuming CE sensitivity. %
However, in this case, the credible intervals on $\mathcal R(z)$ broaden considerably for $z\gtrsim 2.5$.

Although we have chosen a set of representative examples, there are numerous extensions and use cases for this method. While we choose to recover the merger rate as a function of the redshift, we could have chosen the mass distribution, for example. Future work could also include fitting in multiple dimensions instead of just a single dimension. 
For example, we could implement 2D splines to allow for joint constraints on mass distribution vs redshift, or beyond. 
Population analyses of individual events have already been extended to several dimensions~\cite{heinzelNonparametricAnalysisCorrelations2024,heinzelPixelPopHighResolution2024}. 
This feature of our injection and recovery procedure is useful because, given a detected background signal, we could recover real astrophysical distributions. 

Furthermore, multiple interpolation methods can be employed in the model.
This highlights the generality of our algorithm, since different interpolation methods may be better tailored to recover different potential signals. 
For example, linear interpolation will better recover the spectral index of a PL than cubic spline interpolation, which is better suited for a smooth turnover. 
Hence, this algorithm can be used to explore alternative interpolation approaches and evaluate their efficacy to recover different GWB signals in preparation for detection.

This transdimensional spline interpolation model may be extended in several ways. 
We probe a limited number of interpolation methods (linear, cubic splines, Akima splines) in this work--additional interpolation approaches will be better suited to other proposed GWB signals. 
Another extension is to add more general proposals to our RJMCMC to probe the parameter space more efficiently.
For example, a diagonal movement proposal can be added that combines the mechanics of independent horizontal and vertical proposals. 
We expect that this RJMCMC approach, and its future iterations, will play a key role in future GWB detection and characterization, which marks a significant step toward enhancing our signal characterization capabilities in preparation for next-generation GW detectors and anything they may detect.

\acknowledgements
We thank Katerina Chatziioannou for insightful discussions about this work and Jandrie Rodriguez for helping with preliminary investigations for this work.  
This material is based upon work supported by NSF's LIGO Laboratory which is a major facility fully funded by the National Science Foundation (NSF).
The authors are grateful for computational resources provided by the LIGO Laboratory and supported by National Science Foundation Grants No. PHY-0757058 and No. PHY-0823459.
This work was supported by the National Science Foundation Research Experience for Undergraduates program through NSF Grant No. PHY-2150027, the LIGO Laboratory Summer Undergraduate Research Fellowship program, and the California Institute of Technology Student-Faculty Programs.
The computations presented here were conducted partly in the Resnick High Performance Computing Center, a facility supported by Resnick Sustainability Institute at the California Institute of Technology. 

\appendix

\section{\label{app:transdimensional}TRANSDIMENSIONAL FITTING USING RJMCMC}
In this section, we outline the reversible jump Markov chain Monte Carlo approach used throughout the paper. 
\subsection{\label{ssec:transdimensional:rjmcmc_algorithm}RJMCMC Algorithm}
The acceptance probability of moving from model $\boldm$ with $n$ knots activated to model $\boldm'$ with $n'$ knots activated is given by

\begin{widetext}
\begin{align}
    \alpha(\boldm'|\boldm) = \textrm{min}\left[1, \frac{p(\boldm')}{p(\boldm)}\frac{p(\boldd|\boldm')}{p(\boldd|\boldm)} \frac{q(\boldm|\boldm')}{q(\boldm'|\boldm)}|\mathbf{J}|\right],
\end{align}
\end{widetext}
where $p(\boldm)$ is the prior, $p(\boldd | \boldm)$ is the likelihood of the data given the model $\boldm$ [Eq. (\ref{eq:ln_likelihood})], and $q(\boldm| \boldm')$ is the probability of proposing to move from model $\boldm$ to model $\boldm'$. $|\mathbf{J}| = 1$ is the Jacobian between models. 

\subsection{\label{ssec:transdimensional:priors}Priors}
We consider an overall prior $p(\boldm)$. The number of knots that are in use is $n$, the amplitude of those knots is $\bolda_{n}$, and the location of the knots is $\boldx_{n}$. The prior is $p(\bolda_{n}, \boldx_{n}, n)$, which we expand using the product rule
\begin{align}
p(\bolda_{n},  \boldx_n, n) = p(\bolda_{n}, \boldx_n| n)p(n).
\end{align}
The prior probability of our amplitudes and horizontal positions are independent of one another. So we can expand our prior even further:
\begin{align}
    p(\bolda_{n}, \boldx_{n}|n) = \prod_{j=1}^{n} p(a_j|n)p(x_j|n),
\end{align}
where the prior on $\bolda_{n}$ and $\boldx_{n}$ are independent of each other and priors on individual knots are also independent of the prior on other knots.

We place a uniform prior on our amplitudes between points $a_{\textrm{min}}$ and $a_{\textrm{max}}$:
\begin{align}
p(\bolda_{n} | n) = \frac{1}{\left(a_{\textrm{max}} - a_{\textrm{min}}\right)^{n}}
\end{align}
We then place a uniform prior on the location of each of our knots, which we allow to vary around their initial central location by an amount $\Delta x$ in either direction. Therefore,
\begin{align}
    p(\boldx_n | n) = \frac{1}{(2\Delta x)^n}.
\end{align}

Finally, we set a uniform prior on the number of activated control points:
\begin{align}
p(n) = \frac{1}{N - N_{\textrm{min}}},
\end{align}
where $N_{\textrm{min}}$ can be chosen as the minimum number of control points that must always be activated. For linear interpolation or cubic spline interpolation to work $N_{\textrm{min}}=2$, although we discuss how to handle $N_{\textrm{min}}=1$ and $N_{\textrm{min}}=0$ cases below.

It is then straightforward to check that integrating (and summing) this prior over the full range of each parameter gives
\begin{widetext}
\begin{align}
\sum_{n=1}^N\left\{p(n)\left[\prod_j\int dx_j \,p(x_j | n)\right]\left[\prod_{j} \int_{a_{\textrm{min}}}^{a_{\textrm{max}}}\,da_j \;p(a_j|n)\right]\right\}= 1.
\label{eq:prior_integration}
\end{align}
\end{widetext}

\subsection{\label{ssec:transdimensional_proposals}Proposals and acceptance probabilities}
In this subsection we present the different MCMC proposals we use. For each proposal, we calculate the ratio of priors between the ``current'' and ``proposed'' models. We then calculate the ratio of the probability of proposing to move from model $\boldm$ to model $\boldm'$, $q(\boldm|\boldm')$ and the reverse. We present this in terms of a single ratio
\begin{align}
\mathbf{R}(\boldm'| \boldm) = \frac{p(\boldm')}{p(\boldm)}\frac{q(\boldm|\boldm')}{q(\boldm'|\boldm)}.
\end{align}
And so, from the ratios we calculate in this section, the implementation of the acceptance probability is
\begin{align}
    \alpha(\boldm'|\boldm) = \textrm{min}\left[1, \frac{p(\boldd|\boldm')}{p(\boldd|\boldm)} \mathbf{R}(\boldm' |\boldm)|\mathbf{J}|\right].
\end{align}

\subsubsection{Birth proposal}
In this case, we propose to move from a model with $\boldm=(\bolda_{n}, \boldx_n, n)$ to a new model with an extra knot $\boldm'=(\bolda_{n+1}, \boldx_{n+1}, n+1)$. We note that all knot locations $\boldx_n$, are the same in $\boldx_{n+1}$, except for the addition of one extra knot, $x'$. We choose $x'$ by choosing randomly from the $N - n$ points that are currently deactivated. We choose the new grid point $x'_j$ uniformly within $x_\alpha\pm\Delta x$, where $x_\alpha$ is the corresponding center of the underlying grid point. 

We draw $a'$ in two different ways. We draw from its prior a fraction of the time, given by $p_u$. Otherwise, we draw from a Gaussian centered around the amplitude of the point as predicted by the interpolation model $\boldm$, $a_{m}$. The ratio of the proposals is then given by
\begin{align}
    \frac{q(\boldm|\boldm')}{q(\boldm'|\boldm)} = 2\Delta x \frac{p_d}{p_b}\left[\frac{p_u}{a_{\textrm{max}}-a_{\textrm{min}}} - (1-p_u)\frac{e^{-\frac{1}{2}\frac{(a'-a_m)^2}{2\sigma^2}}}{\sqrt{2\pi\sigma^2}}\right]^{-1},
\end{align}
where $\sigma$ is a scale parameter chosen depending on the problem, $p_b$ is the fraction of all proposals that are birth proposals and $p_d$ is the fraction of all proposals that are death proposals.  In this work we take $\sigma=1$ throughout, although for very strong injections choosing a smaller $\sigma$ could increase the birth acceptance rate. Taking into account the relative prior volumes, we have
\begin{align}
    \mathbf{R}(\boldm'|\boldm) = \frac{p_d}{p_b}\left[p_u - (1-p_u)\frac{(a_{\textrm{max}}-a_{\textrm{min}})e^{-\frac{1}{2}\frac{(a'-a_m)^2}{2\sigma^2}}}{\sqrt{2\pi\sigma^2}}\right]^{-1}.
\end{align}

\subsubsection{Death proposal}
The death proposal is conceptually similar to the birth proposal. We propose to move from the current model, $\boldm = (\bolda_{n}, \boldx_n, n)$ to a new model $\boldm'=(\bolda_{n-1}, \boldx_{n-1}, n-1)$. We randomly deactivate one of the $n$ activated points, $x'$. The probability of moving from $\boldm$ to $\boldm'$ corresponds to the probability that we chose that specific $x'$. Likewise, the reverse probability of moving from $\boldm'$ back to $\boldm$ is the probability that we have chosen to reactivate $x'$ from a choice of $N-n+1$ available deactivated points, multiplied by the probability of choosing $a'$ and $x'$. This must take into account both of the methods for choosing a new point that we discussed in the previous section. Putting these things together, we find
\begin{align}
    \mathbf{R}(\boldm'|\boldm) = \frac{p_b}{p_d}\left[p_u - (1-p_u)\frac{(a_{\textrm{max}}-a_{\textrm{min}})e^{-\frac{1}{2}\frac{(a'-a_m')^2}{2\sigma^2}}}{\sqrt{2\pi\sigma^2}}\right],
\end{align}
where the new notation $a_{m'}$ is the \textit{prediction} for the amplitude at point $x'$ using the interpolated model $\boldm'$.

\subsubsection{Other proposals}
We consider three other MCMC proposals, all of which have $\mathbf{R}=1$.
\begin{enumerate}
    \item Redraw the amplitude of a randomly chosen knot from its prior.
    \item Redraw the amplitude of a randomly chosen knot from a Gaussian centered around the current point. The scale of that Gaussian is set depending on the problem, and can be randomized to allow for different sized jumps. In this work, we randomly chose a small-scale factor uniformly between 0.1 and 0.01. 
    \item Redraw the $x$ value of a randomly chosen knot within its valid prior region.
\end{enumerate}

\section{BAYES FACTORS AND ODDS RATIOS}
\label{apx:bayes}
By allowing the dimension of the interpolation model to vary, we naturally probe a range of distinct models, where each configuration conceivably represents a different model. 
The final set of posterior samples, when properly converged, represents a posterior across those models. We can use these samples to compare the preference for different models. 
However, the transdimensional nature of the comparison leads to irreducible ambiguities when defining a metric for model selection--e.g., Bayes' factors. We elaborate on this here, and provide suggestions for Bayes' factor calibration. %

Consider fitting $\Omegagw(f)$; one could construct an odds ratio between a signal model ($n>0$) and a noise model ($n=0$) as follows:
\begin{align}
    O_{\mathcal N}^{S} = \frac{p(n>0|d)}{p(n=0|d)} \approx \frac{\textrm{No. of samples with }n>0}{\textrm{No. of samples with }n=0}\,.\label{eq:odds}
\end{align}
If it is equally likely that the number of activated knots range from $n=0$ to $n=N_{\textrm{max}}$, then $p(n=k) = 1/(N_{\textrm{max}}+1)$. This means that the prior odds for the signal model to the noise model is given by:
\begin{align}
    \frac{p(n>0)}{p(n=0)} = \frac{N_{\textrm{max}}/(N_{\textrm{max}}+1)}{1/(N_{\textrm max}+1)} = N_{\textrm{max}}.
\end{align}

In this case we have designed prior odds that favor the presence of a signal compared to noise. This is not a typical choice in the context of signal detection, where it is often more preferable to have false negative results compared to false positive results. A statistic with equal prior odds on signal and noise (a Bayes factor) is more common and is given by
\begin{align}
    B_{\mathcal N}^S = \frac{1}{N_{\textrm{max}}} O_{\mathcal N}^S \approx \frac{1}{N_{\textrm{max}}}\frac{\textrm{No. of samples with }n>0}{\textrm{No. of samples with }n=0} .
\end{align}.
Conversely, the interpolation model with large $N_\textrm{max}$ has a much larger prior volume than the typical PL model that is used. This is not a problem if the extra parameters fill their prior distributions. However in stochastic analyses like those presented here, very broad uninformative priors are often used, such that the data produce upper limits well below the prior that was set. This can result in unintended outcomes.

To explore the push and pull between the prior odds and the Occam penalty for our specific case more concretely, we consider a representative toy example that captures the spirit of the problem. 
We place a uniform prior on the number of knots $n$. If we consider the evidence for the signal model only, then we use $p(n) = 1/N_{\textrm{max}}$. 
The evidence for the signal model is then
\begin{align} \label{eqn:Zs}
    \mathcal Z_{S} = \sum_{n=1}^{N_{max}} &\int p(d | a, n) p(a, n)da\,,\\
    = \sum_{n=1}^{N_{max}}& \left[\frac{1}{N_{\textrm{max}}} \frac{1}{(a_{max} - a_{min})^n}\right.\\ &\times \left.\prod_{i}\int_{a_{min}}^{a_{max}} \mathcal L_{N} \Theta(a_i - a_{crit}) da_i\right]\,.
\end{align}
In Eq. ~\eqref{eqn:Zs}, we assume the amplitudes at different nodes are independent (likely a faulty assumption in the end, but this is a toy example) and that the likelihood is roughly constant and equal to $\mathcal L_{N}$, up to some value $a_{crit}$ before sharply falling to zero. This is the typical situation for GWB analyses upper limits are set on the GWB can be set as a function of frequency. We model this with the Heaviside function $\Theta$. The priors on $a$ and $n$ have been pulled out. Defining $\xi = (a_{crit} - a_{min}) / (a_{max} - a_{min})$, the ratio between the size of the posterior and the size of the prior, we find:
\begin{align}
    \mathcal Z_{S} &= \frac{\mathcal L_{N}}{N_{\textrm{max}}} \sum_{n=1}^{N_{max}}\xi^n\,,\\
    &=\frac{\mathcal L_N}{N_{\textrm{max}}} \xi\frac{1 - \xi^{N_{\textrm{max}}}}{1 - \xi} \textrm{ assuming $\xi<1$}.
\end{align}
The odds ratio for signal to noise, then, is given by
\begin{align}
    \mathcal O_{\mathcal N}^S = \frac{\mathcal Z_S}{\mathcal Z_{\mathcal N}} \frac{p(n>0)}{p(n=0)} \approx \xi\frac{1 - \xi^{N_{\textrm max}}}{1 - \xi}.
\end{align}

\begin{figure}
    \centering
    \includegraphics[width=\linewidth]{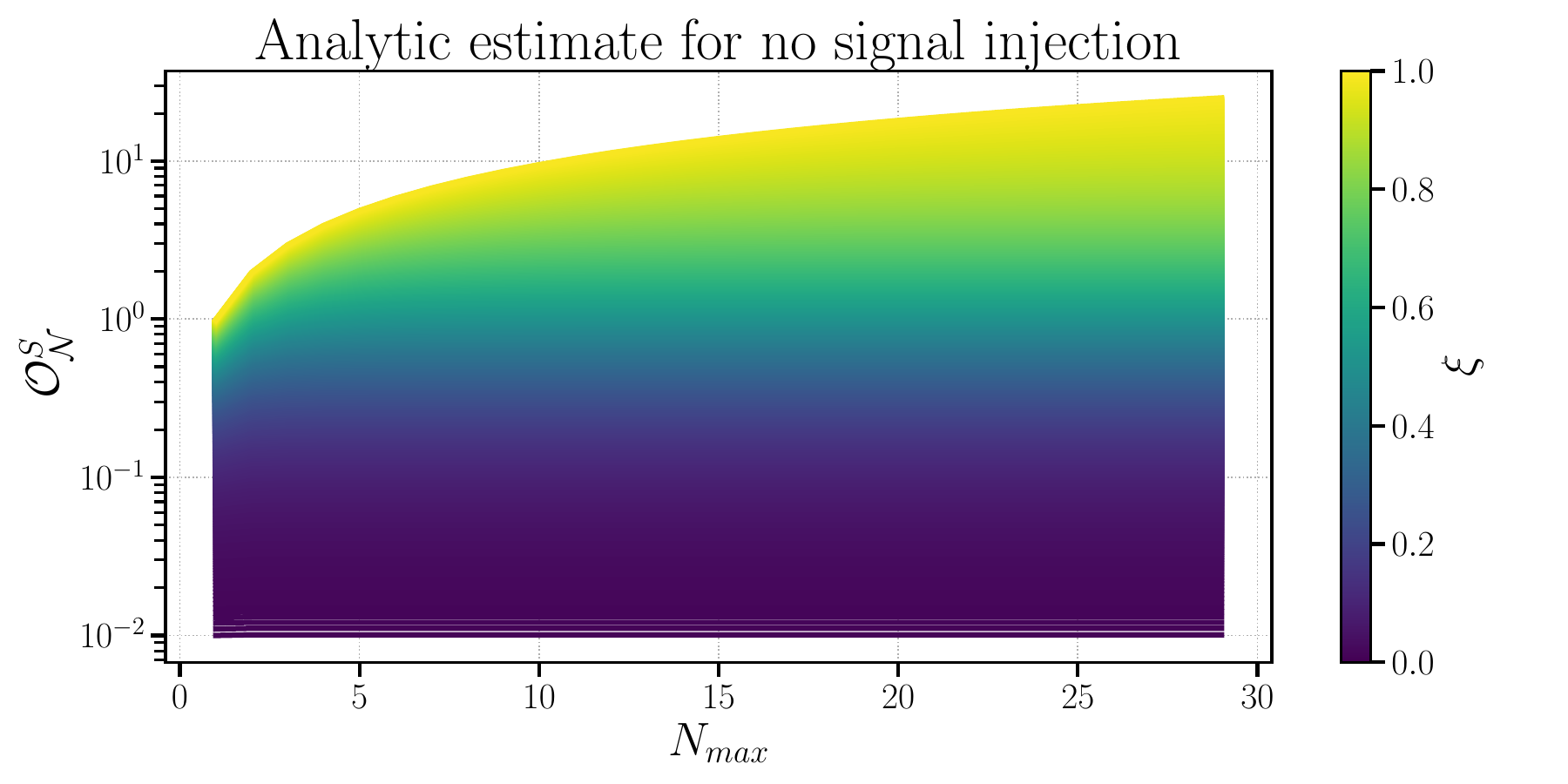}
    \caption{We show $\mathcal{O}^{S}_{\mathcal N}$ for our toy model when no signal is present, for different choices of $N_{\textrm{max}}$ (x axis) and $\xi$ (colors). We see that for $\xi\approx 1$ the odds grow with $N_{\rm max}$. But for $\xi \lesssim 0.5$ they grow slower and return odds that are indifferent to either model.}
    \label{fig:analytic_odds_estimate_noise}
\end{figure}

We show how this scales for different values of $\xi$ (colors) and $N_{\textrm{max}}$ in Fig.~\ref{fig:analytic_odds_estimate_noise}.  We can see that for $\xi \approx 1$, there is essentially no Occam penalty for adding a spline knot, and so the odds ratio even when no signal is present, can grow with $N_{\textrm{max}}$. However, for $\xi \lesssim 0.5$, the odds ratio does not grow significantly with $N_{\rm max}$.
This leads to an unavoidable ambiguity of our model: when  $\xi$ is small, the prior odds offset the very large Occam penalty incurred by allowing for complex models, while when $\xi \approx 1$ the odds grow with $N_{\textrm{max}}$. 

This implies that the transdimensional approach should be handled with care in the context of signal detection. When claiming a detection with a threshold significance, we expect low detectable complexity in our model, and so we advise choosing a small $N_{\textrm{max}}$, e.g. $\lesssim 5$. This still allows us to potentially be more sensitive to e.g. a PL that breaks in our sensitive band, but keep the detection power of a simple PL model. In this framework, frequentist statistics such as the SNR of Fig.~\ref{fig:cumulative_SNR} should provide a useful guideline.

If the noise is well understood, one can employ a simulation-based approach to calibrate the odds ratio. This would involve first selecting an $N_{\rm max}$ and then creating many simulated frequency-domain datasets with $\Omegagw(f)=0$ and a variance $\sigma^2(f)$ and calculating a distribution on $\mathcal O^{S}_{N}$. One can then calculate a false alarm rate for $\mathcal O^{S}_{N}$ on the real data, under this noise model. This would eliminate the ambiguity introduced into the odds ratio (and therefore Bayes factors) due to the choices in maximum model complexity and prior bounds.

Finally, one can also calculate Bayes factors by estimating the evidence for signal $\mathcal Z_{S}$, and noise $\mathcal Z_{\mathcal N}$, directly using nested sampling~\cite{skilling_nested_2006} or thermodynamic integration~\cite{gelmanSimulatingNormalizingConstants1998,lartillotComputingBayesFactors2006a}. This allows for arbitrarily high Bayes factors and odds ratios to be calculated, while the RJMCMC technique is limited to odds ratios as large as the number of MCMC samples used. This is the method employed by BayesWave~\cite{Cornish:2014kda,Cornish:2020dwh}.

\section{LINEAR VS SPLINE AND AKIMA INTERPOLATION} \label{apx:interp}

\begin{figure*}[]
    \includegraphics[width=0.32\textwidth,clip=true]{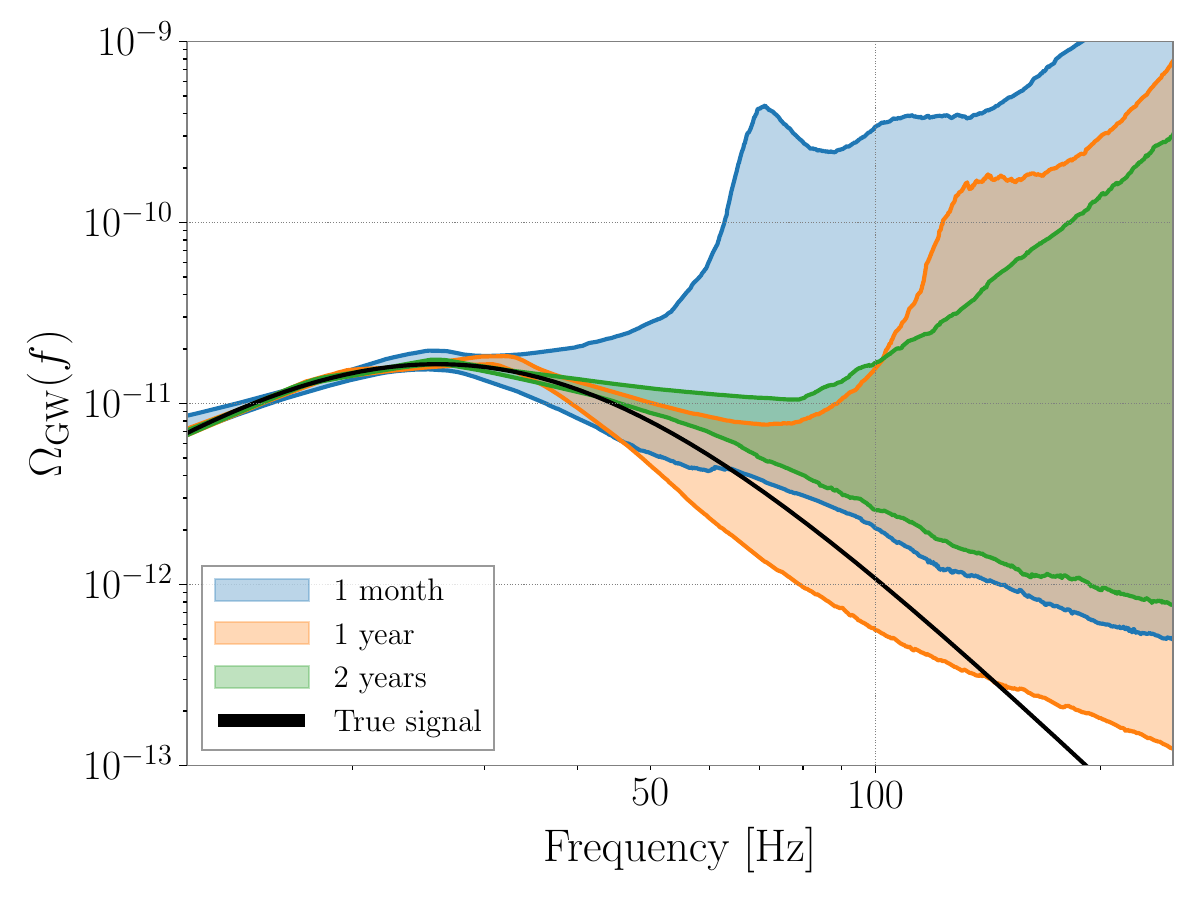}
    \includegraphics[width=0.32\textwidth,clip=true]{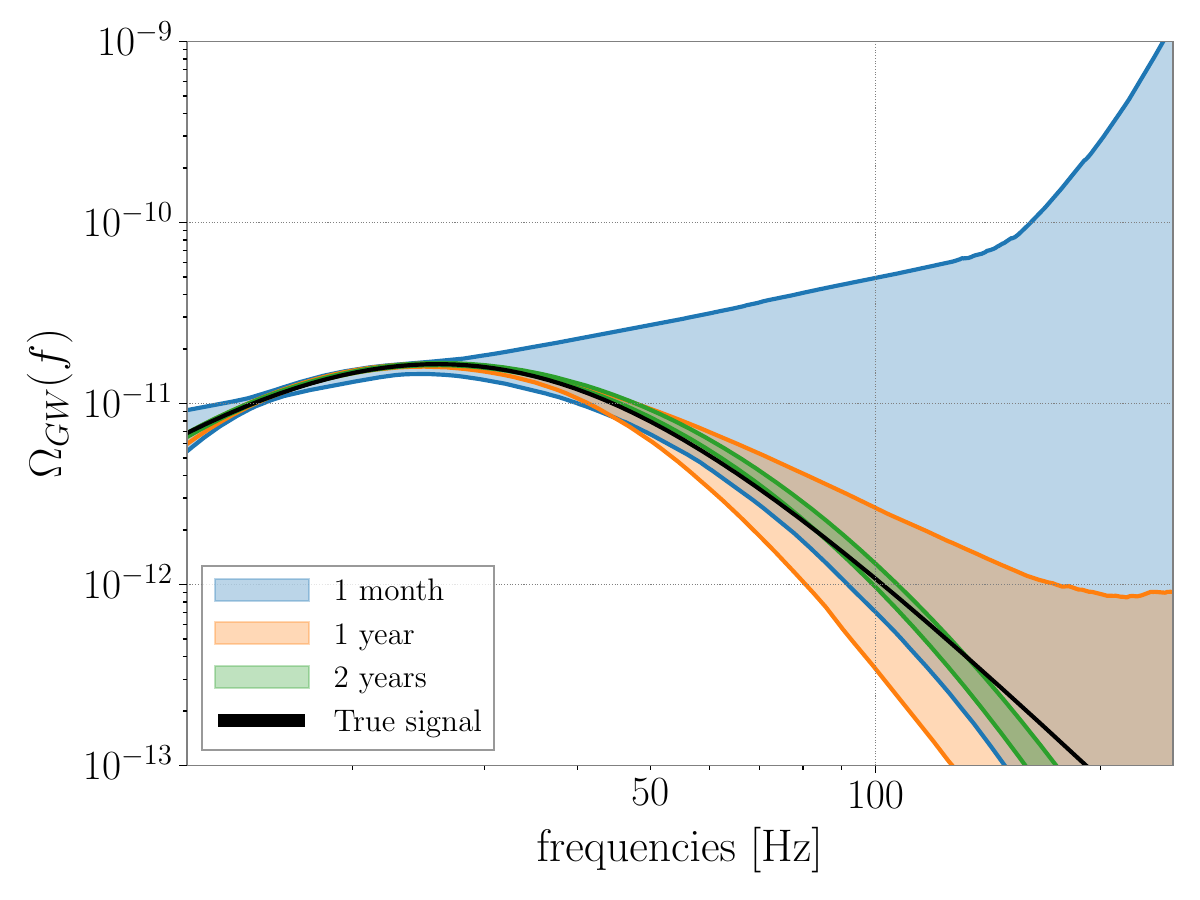}
    \includegraphics[width=0.32\textwidth,clip=true]{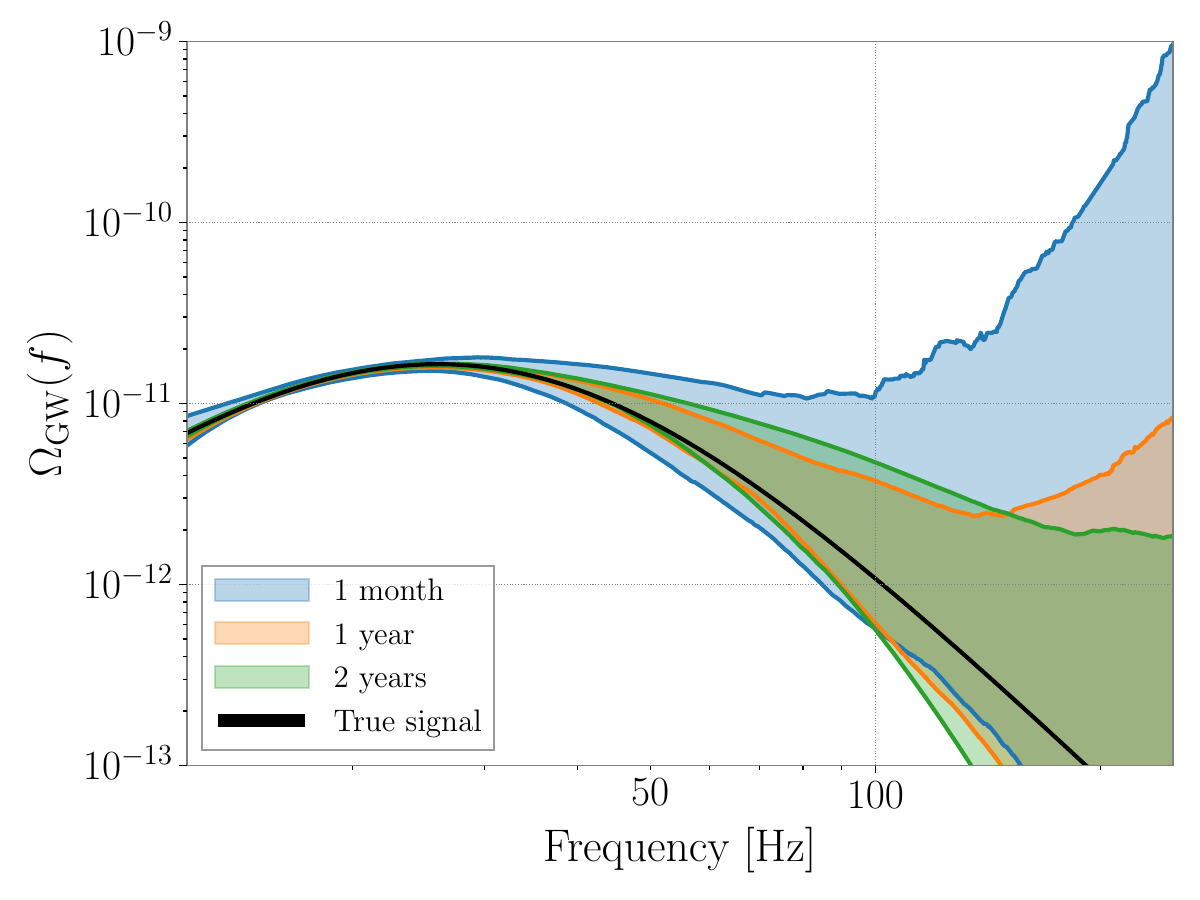}\\
     \caption{ \label{fig:FOPT_intrp_compare} 
     Results for the recovered posteriors from injecting a FOPT-informed smoothed BPL with three different interpolation methods with varying observation tines and CE sensitivity. We show the 95\% confidence recovery envelopes for linear (left), cubic spline (middle), and Akima spline (right) interpolation.
     As the observing time increases, the 95\% confidence recovery envelopes for each interpolation type converges to the injected signal (red dashed line).
     }
\end{figure*}

\begin{figure*}[]
    \includegraphics[width=0.32\textwidth,clip=true]{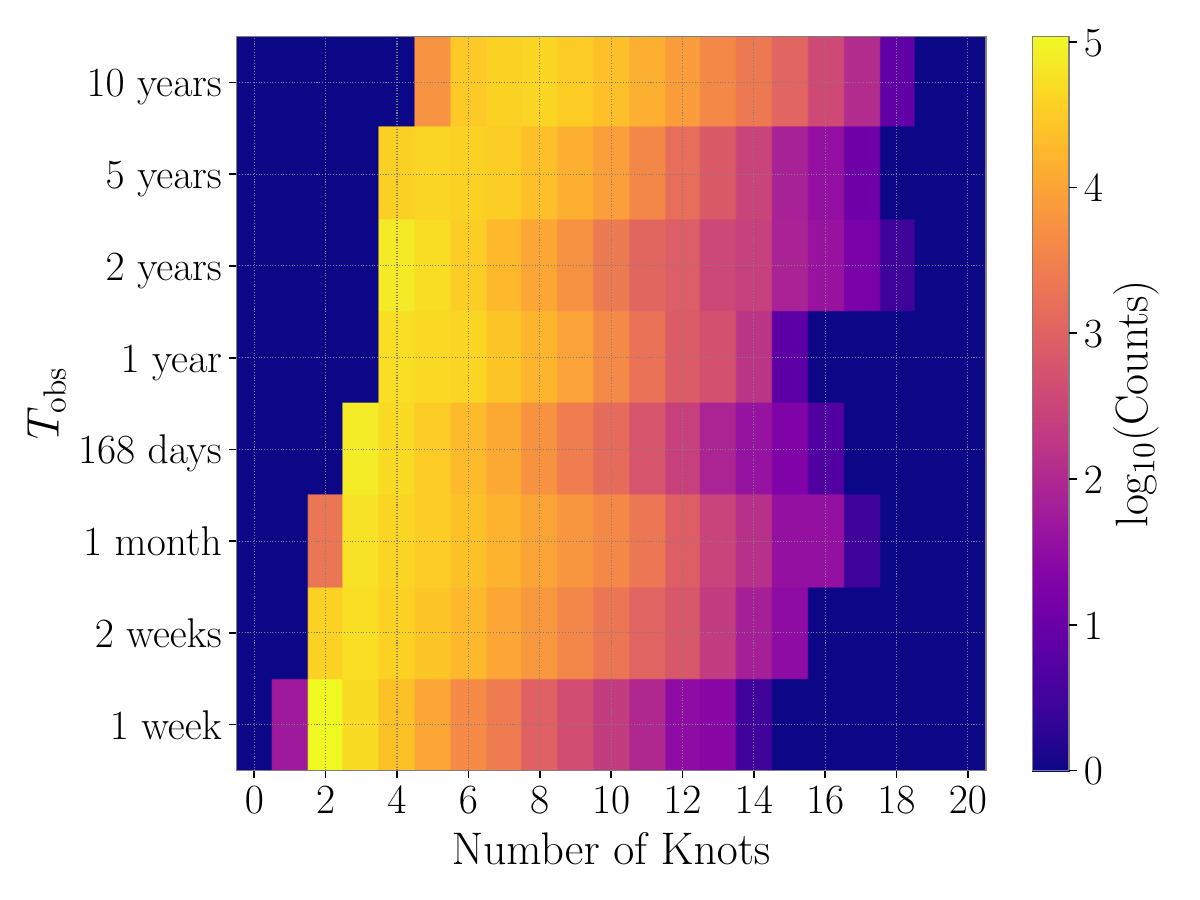}
    \includegraphics[width=0.32\textwidth,clip=true]{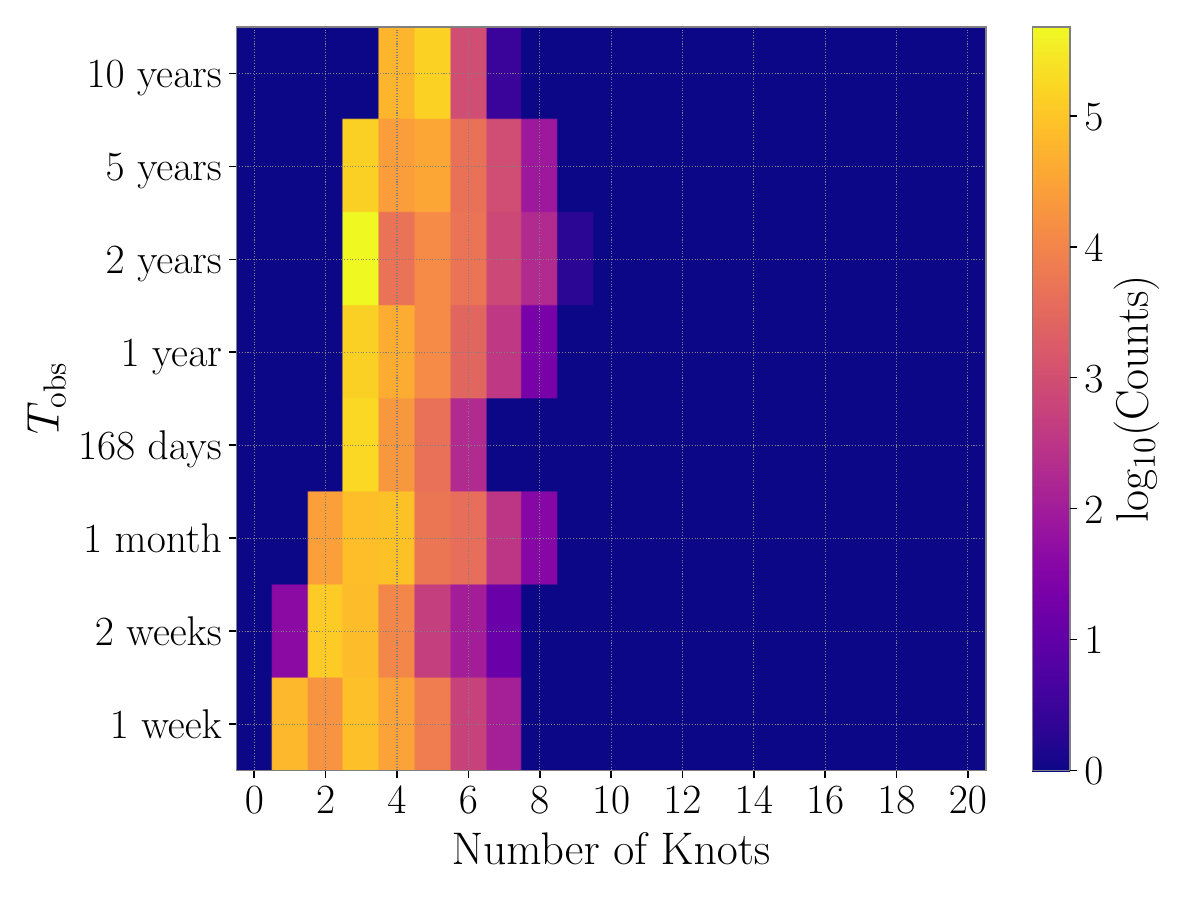}
    \includegraphics[width=0.32\textwidth,clip=true]{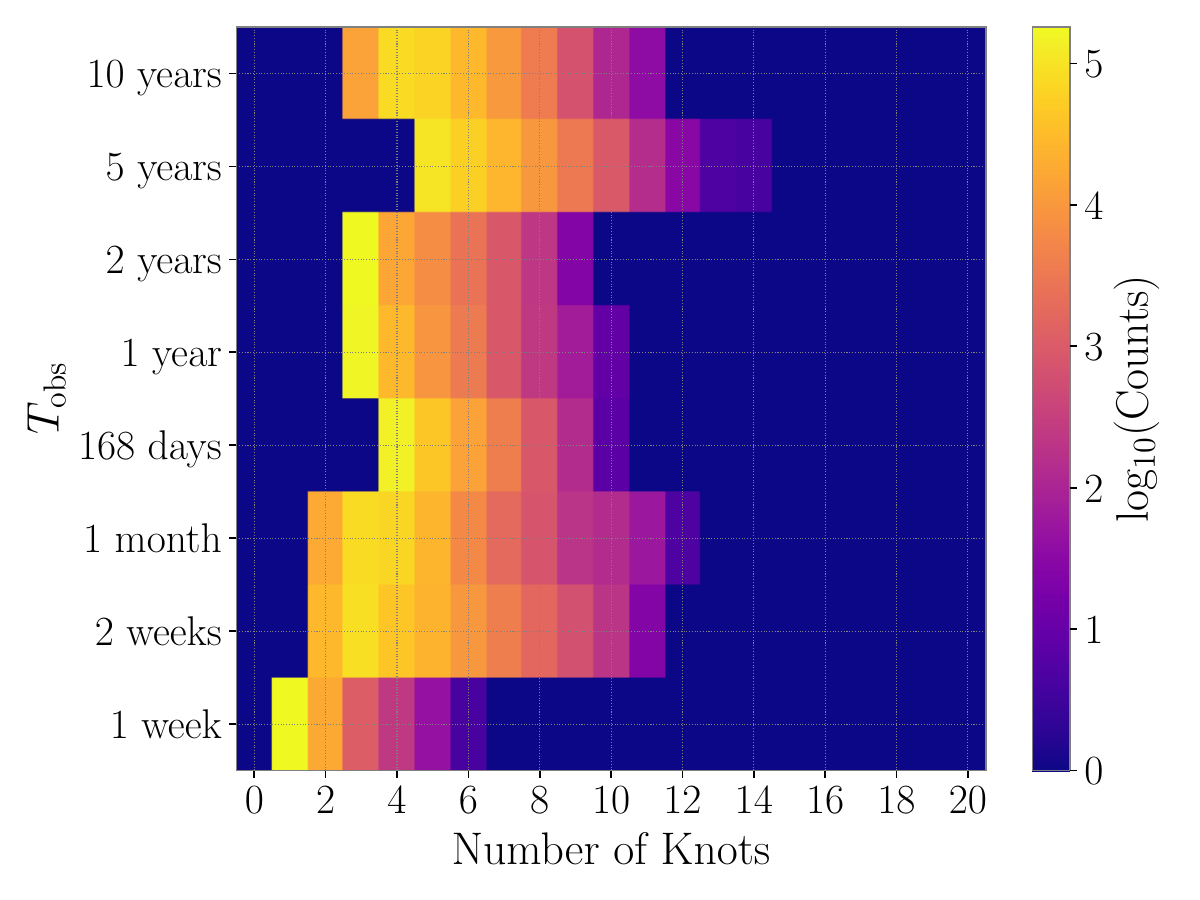}\\
    \caption{ \label{fig:FOPT_intrp_compare_knots} 
    Results for the number of knots corresponding to the posterior recovery envelopes shown in Fig. \ref{fig:FOPT_intrp_compare}. We show a heat map of the number of knots per posterior for linear (left), cubic spline (middle), and Akima spline (right) interpolation. 
    Linear interpolation requires a greater number of knots to fit a smooth curve while the cubic and Akima splines require predominantly between two and three knots for each posterior. 
    }
\end{figure*}

\begin{figure}[]
    \includegraphics[width=\columnwidth,clip=true]{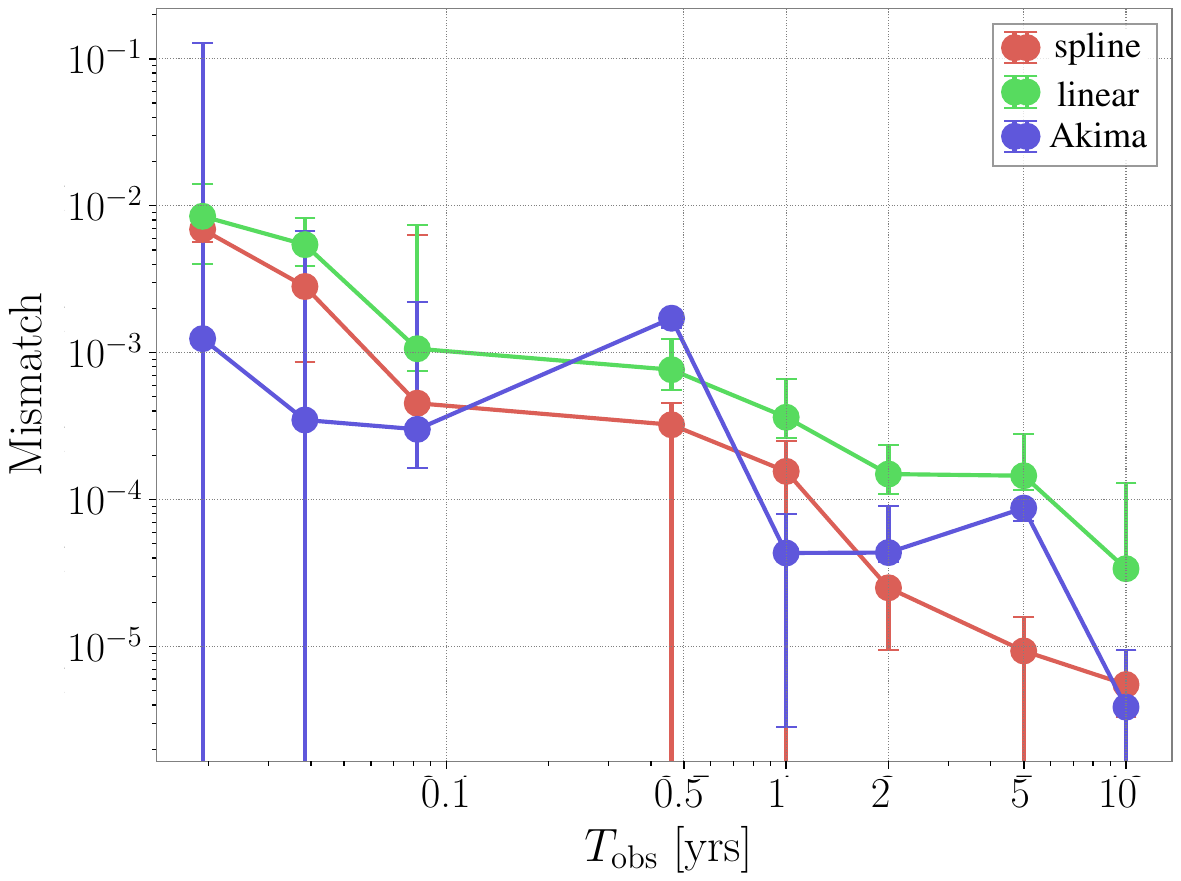}\\
     \caption{ \label{fig:FOPT_mismatch} 
     Mismatch between the injected signals and recovered posteriors as a function of observing time for each interpolation technique. 
     The markers correspond to the mismatch with the average posterior behavior.
     The error bars correspond to the mismatches with the signal and the upper and lower boundaries of the 95\% confidence recovery envelopes from Fig. \ref{fig:FOPT_intrp_compare}.
     The mismatches generally decrease as a function of observing time and cubic spline interpolation (red) most consistently produces the best fits for an injection.}
\end{figure}

In this appendix, we compare the different types of interpolation used in this work: linear, cubic spline, and Akima. We refer back to Sec. \ref{sec:intrp_models} for specifics on each interpolation type. 

We begin by comparing the recoveries of an injected smoothed BPL signal (see Sec. \ref{sec:fopt}). 
Figure \ref{fig:FOPT_intrp_compare} presents the 95\% recovery envelopes for various observation times for linear (left), cubic spline (middle), and Akima spline (right) interpolation. 
All three approaches can resolve the low-frequency segment of the signal well.  
However, of the three options, cubic spline interpolation resolves higher frequencies best where there is less detector sensitivity. 
Therefore, cubic splines are useful if we choose to probe all frequency regimes of a signal recovery. 

The other two interpolation approaches also have their advantages.
Linear interpolation will better discern whether there is strict PL behavior. 
Section \ref{sec:subtraction} explores an example of linear interpolation, where we consider a background subtraction signal. 
On the other hand, Akima spline interpolation seems to recover both PL and non-PL behavior. 
For small observation times (blue and orange), the 95\% confidence posterior envelope generally appears to recover a PL. 
As the observation time increases, the Akima spline interpolation yields a smooth turnover that matches the injected signal. 
The limitations to using Akima spline interpolation are that there are broader posterior envelopes than for cubic spline interpolation for the BPL. However, this is likely due to cubic splines being able to describe the specific injected model with fewer parameters because a BPL is well described by cubic splines.

We now consider the number of knots required by each approach to recover the injected signal as a function of observation time, shown in Fig. \ref{fig:FOPT_intrp_compare_knots}.
Linear interpolation requires the greatest number of knots for all observation times (left).
This makes sense because we are trying to fit a smooth and differentiable injection with linear line segments. 
The cubic (middle) and Akima (right) spline interpolation methods both require mostly two or three knots in their posteriors. 
This means fewer knots and a more consistent distribution of the number of knots than in the linear interpolation approach. 

Finally, we calculate a statistic to measure the precision with which each fitting method reconstructs the injected signal.
Motivated by LVK matched-filtering searches, we utilize the mismatch \cite{Blackman_2017}
\begin{equation} \label{eqn:mismatch}
    M(d, s) = 1 - \frac{\langle d,s\rangle}{\sqrt{\langle d,d\rangle\langle s,s\rangle}}\,
\end{equation}
where $d$ is in the injected signal and $s$ is the recovered signal. 
Following the conventions from Sec. \ref{sec:search_methods_overview}, we define the inner product from Eq.~\eqref{eqn:mismatch} in terms of our GWB model, $\Omega_{\mathcal{M}}$, and the injected signal, $\hat{C}$:
\begin{equation}
    \langle \Omega_{\mathcal{M}} , \hat{C}\rangle = 4 {\rm Re} \int^\infty _{-\infty} \frac{\Omega_{\mathcal{M}}(f) \hat{C}^*(f)}{\sigma^2(f)}df\,.
\end{equation}
We weigh the inner product by the variance, $\sigma^2(f)$, associated with the injection. 
Generally, $M(\Omega_{\mathcal{M}} , \hat{C}) \leq 1$, where $M(\Omega_{\mathcal{M}} , \hat{C}) = 0$ implies that two signals are identical.

We calculate the mismatches between the boundaries of the 95\% confidence envelopes and the average posterior and plot the results in Fig. \ref{fig:FOPT_mismatch}. 
Right away, we notice that the mismatches decrease as a function of the observation time for each of the interpolation methods. 
The mismatches for all three interpolation methods improve by almost 3 orders of magnitude between 1 week of observing time and 10 years of observing time. 
This trend indicates that the fits for each of the interpolation methods are improving as we increase sensitivity through longer observing times, which is what we expected. 

The mismatches can be utilized to quantify the best approach for interpolating. 
Because a lower mismatch correlates to a better recovery of the posterior, the interpolation method with the lowest mismatch is the best option for a particular amount of observing time. 
The mismatches for cubic splines (red) and linear interpolation (green) monotonically decrease as a function of the observing time, where the cubic splines have mismatches consistently smaller than those of linear interpolation. 
However, the Akima spline mismatches (blue) do not behave monotonically.
For some observation times, Akima splines recover the correct shape of the injection better than cubic splines. 
At certain observing times, like $T_{\rm obs} = 168$ days or $T_{\rm obs} = 5$ years, cubic splines and linear interpolation recover the signal better than Akima splines.
Although Akima splines are the best interpolation method for recovering a signal for most observation times, cubic splines consistently recover injections well, making cubic splines a more reliable interpolation option. 

We conclude from this investigation that each interpolation choice can recover an injected signal and has optimal implementations. 
Linear interpolation is the best option for recovering a PL signal, whereas cubic or Akima splines are more effective in recovering a smooth, differentiable signal.
Linear interpolation uses more knots, which risks overfitting. 
However, cubic and Akima splines require fewer knots in their posteriors to recover the signal, which reduces the potential for overfitting. 

In the specific instance of the FOPT smoothed BPL, cubic splines are the optimal interpolation choice. Therefore, we employ cubic splines in Sec. \ref{sec:fopt} of the main text. In contrast, linear splines are more useful for CBC-informed injection (Sec. \ref{sec:CBC}) and subtraction injection (Sec. \ref{sec:subtraction}). By implementing various interpolation techniques within the main body of the paper, we highlight the respective advantages of each method, which are further elaborated upon in this appendix.

\bibliography{apssamp}

\end{document}